\documentclass[aps,prd,eqsecnum,superscriptaddress,twocolumn,nofootinbib,floatfix,preprintnumbers]{revtex4-1}

\usepackage{graphicx}
\usepackage{amssymb}
\usepackage{amsmath}
\usepackage{color}
\usepackage{multirow}

\begin{document}

\newcommand*{\TTU}{Department of Physics and Astronomy, Texas Tech University, Lubbock, TX 79409-1051 (USA)}\affiliation{\TTU}
\newcommand*{\URI}{Department of Physics, University of Rhode Island, Kingston, Rhode Island 02881 (USA)}\affiliation{\URI}
\title{Multi-waveform cross-correlation search method for intermediate-duration gravitational waves from gamma-ray bursts}

\author{Eric Sowell}\email{eric.sowell@ttu.edu}\affiliation{\TTU}
\author{Alessandra Corsi}\affiliation{\TTU}
\author{Robert Coyne}\affiliation{\URI}
\date[\relax]{compiled \today }


\begin{abstract}
Gamma-ray Bursts (GRBs) are flashes of $\gamma$-rays thought to originate from rare forms of massive star collapse (long GRBs), or from mergers of compact binaries (short GRBs) containing at least one neutron star (NS). The nature of the post-explosion / post-merger remnant (NS versus black hole, BH) remains highly debated. In $\sim 50\%$ of both long and short GRBs, the temporal evolution of the X-ray afterglow that follows the flash of $\gamma$-rays is observed to ``plateau'' on timescales of $\sim 10^2-10^4$\,s since explosion, possibly signaling the presence of energy injection from a long-lived, highly magnetized NS (magnetar). The Cross-Correlation Algorithm (CoCoA) proposed by [R. Coyne et. al., Phys Rev D. \textbf{93} 104059 (2016)] aims to optimize searches for intermediate-duration ($10^2-10^4$\,s) gravitational waves (GWs) from GRB remnants. In this work, we test CoCoA on real data collected with ground-based GW detectors. We further develop the detection statistics on which CoCoA is based to allow for multi-waveform searches spanning a physically-motivated parameter space, so as to account for uncertainties in the physical properties of GRB remnants.

\end{abstract}

\flushbottom
\maketitle

\thispagestyle{empty}

\section{Introduction}
\label{sec:intro}
Gamma-ray Bursts (GRBs) are the most relativistic explosions we know of in the universe. Observationally, they are characterized by a burst of $\gamma$-rays followed by a slower-evolving, multi-wavelength emission dubbed ``afterglow''. They are divided in two major classes based on the duration of their $\gamma$-ray emission \citep{Kouveliotou1993}.
Long-duration GRBs, whose $\gamma$-ray emission lasts for more than 2\,s, are thought to originate from rare forms of massive star collapses. On the other hand, short GRBs with duration less than 2\,s are linked to mergers of compact binaries containing at least one neutron star (NS). The nature of the GRB central engine, also referred to as GRB remnant, is still highly debated as its properties cannot be probed directly using light. While it had been theorized that black holes (BHs) may act as central engines of both short and long GRBs \cite{1993Woosley,1999Popham,1999Piran,2013Lei,2017Lei}, the identification of ``plateaus'' in $\sim 50$\% of both short- and long-duration GRBs observed by \textit{Swift} (e.g., \cite{Dainotti2017,Kisaka2017}) has renewed interest in the role of long-lived highly-magnetized neutron stars (magnetars) as GRB central engines \cite{Nousek2006,Zhang2006,Liang2007,Starling2008,Bernardini2012,Gompertz2013,Rowlinson2013,Yi2014,2014Ravi,2017Yu}. 

The recent detection of gravitational waves (GWs) from the in-spiral phase of a compact binary merger (GW170817) associated with the short GRB\,170817A \cite{GW170817Discovery} has spurred new investigations into the nature of GRB remnants \cite{2017Postmerger, 2019Postmerger}.
 Some models predict that magnetars formed in GRB explosions may undergo deformations, such as magnetic field induced ellipticities \cite{1996Bonazzola,2001Palomba,2002Cutler}, unstable bar-modes \cite{LaiShapiro1995,Corsi2009}, and unstable r-modes \cite{1998Owen,1998Lindblom,1998Andersson}, that would make them efficient GW emitters.  A detection of GWs in coincidence with a GRB X-ray plateau would provide clear evidence that a magnetar can act as a GRB central engine \citep[e.g.,][]{Zhang2001,Corsi2009}.

The Cross-Correlation Algorithm (CoCoA) proposed by Coyne et al. \cite{2016Coyne} is a GW data analysis technique that aims to optimize searches for intermediate-duration ($\sim 10^3$\,s) GWs from GRB remnants. While several other methods have been used for this purpose \cite[e.g.,][]{2017Postmerger,Thrane2011,Abbott2019,2019Postmerger,2018Sun,2018Miller, 2016Klimenko, Oliver2019}, CoCoA is among a small number of methods (such as \cite{2016Klimenko}) that can target both slow- and fast-evolving signals (${\cal O}(10)\,{\rm mHz\,s}^{-1} \lesssim \dot{f}_{\rm{max}} \lesssim {\cal O}(10)\,{\rm Hz\,s}^{-1}$) while using a technique that bridges stochastic and continuous wave searches \cite{2016Coyne}.
Indeed, as shown by Coyne et al. \cite{2016Coyne}, the strength of CoCoA lies in its tuneability for sensitivity and robustness. Traditional in-spiral and continuous wave GW searches make use of matched filters that maximize sensitivity at the expense of robustness, thus requiring highly accurate GW waveforms \citep{Schutz1991,Jar1998,Hinder2013,Apostolatos1995,Samson2014}. 
At the other extreme, stochastic (based on cross-correlating the data of two different GW detectors) and burst (based on excess power) searches maximize robustness at the expense of sensitivity \cite{Brady2000,Allen2005,Cutler2005,Abbott2007d,Abbott2009b,Coughlin2011,Abadie2014}. CoCoA allows one to smoothly tune search robustness and sensitivity in between these two extremes. 

Here, we further develop the CoCoA algorithm so as to make it a practical tool for real GW data analyses. Specifically, we (i) adapt the pipeline so that it can handle real data from the Laser Interferometer Gravitational Wave Observatory (LIGO) and Virgo (rather than simulated Gaussian data only, as in \cite{2016Coyne}); (ii) we re-work the cross-correlation detection statistic on which CoCoA is based so that the algorithm can be employed to carry out multi-waveform searches spanning a realistic parameter space (as opposed to only single-waveform analyses); (iii) we make more realistic estimates of the detection efficiency by including uncertainties in the delay between the GRB trigger time and the start of the GW signal, and by accounting for non-ideal sky locations. 

This paper is organized as follows. In Section \ref{Sec:II} we briefly review CoCoA as developed by Coyne et al. \cite{2016Coyne}. In Section \ref{Sec:Waveforms} we describe the waveforms on which the performance of CoCoA is tested. In Section \ref{Sec:Noise} we compare results of searches run over real noise to results when simulated data are used.  In Section \ref{Sec: Template} we introduce the CoCoA multi-trial statistic for spanning a broad parameter space. In Section \ref{sec:bank} we test CoCoA's multi-trial statistics and quantify its sensitivity and detection efficiency for searches of secularly-unstable magnetars.  Finally, in Section \ref{sec:conclusion} we summarize our results and conclude.

\section{The Cross-Correlation Algorithm (C\lowercase{o}C\lowercase{o}A)}
\label{Sec:II}
The detection of GWs that last for durations in the range $10^2 - 10^4$\,s requires different data analysis techniques than those used in traditional inspiral/continuous wave searches. If the waveform of the GW signal can be accurately predicted, then matched filtering is the ideal technique as it maximizes sensitivity \cite{Schutz1991,Jar1998}. On the other hand, if the predicted GW signal is affected by large uncertainties, more robust data analysis techniques are necessary. One of these is the so-called ``stochastic'' method, which requires no prior knowledge of the evolution of the GW signal and is based on cross-correlating the output of two different detectors, under the assumption that the noise of the two detectors is uncorrelated. 

The cross-correlation method first developed by Dhurandar et al. \cite{Dhurandhar2008} for continuous GW searches, and later adapted by Coyne et al. \cite{2016Coyne} to searches of intermediate-duration signals, targets quasi-monochromatic GWs whose time-frequency evolution is known to a certain degree. The resulting (single-trial) semi-coherent statistic bridges the gap between matched-filtering (i.e., fully-coherent) and stochastic-like methods, allowing one to tune the search sensitivity and robustness in between the two extremes of most sensitive but least robust, and least sensitive but most robust. 
In this Section, we briefly review the (single-trial) cross-correlation statistic following closely the notation adopted by \cite{2016Coyne}. 

\subsection{The cross-correlation statistic}
At any given time $t$, a GW detector output $x(t)$ can be represented as the linear combination of a GW signal, $h(t)$, and noise, $n(t)$:
\begin{equation}
x(t) = h(t) + n(t).
\end{equation}
Spectral information about the detector output $x(t)$ can be obtained by performing a Discrete Fourier Transform (DFT) on each of $N_{\rm{SFT}}$ data segments of identical duration $\Delta T_{\rm{SFT}}$ (Short Fourier Transform; SFT) \cite{Dhurandhar2008}: 
\begin{equation}
\label{eq:FFTnoW}
\tilde{x}_{I}[f_{k}]=\frac{1}{f_{\rm{s}}}\sum\limits_{{l=0}}^{N_{\rm{bin}}-1}x[t_{l}]e^{-2\pi i f_{k}(t_{l}-T_{i}+\Delta T_{\rm{SFT}}/2)}.
\end{equation}
\begin{equation}
\label{eq:FFTW}
\tilde{x}_{I}[f_{k}]=\sum\limits_{{l=0}}^{N_{\rm{bin}}-1}w[t_{l}]x[t_{l}]e^{-2\pi i f_{k}(t_{l}-T_{i}+\Delta T_{\rm{SFT}}/2)}.
\end{equation}
In the above Equation \ref{eq:FFTW}, $w[t_{l}]$ is a windowing function applied to reduce spectral leakage\footnote{In Eq. \ref{eq:FFTW} the windowing function, $w[t_{l}]$, absorbs the factor 1/$f_{\rm{s}}$}; $N_{\rm{bin}}$ refers to the number of frequency bins within each SFT, defined as $N_{\rm{bin}} = \Delta T_{\rm{SFT}} \times f_{s}$ where $f_{s}$ is the sampling frequency; $f_{k}$ is the frequency corresponding to the $k$-th frequency bin:
\begin{equation}
f_{k} = \frac{k}{\Delta T_{\rm{SFT}}} \quad \rm{for} \quad k = 0,...,N_{\rm{bin}}/2-1
\end{equation}
\begin{equation}
f_{k} = \frac{k - N_{\rm{bin}}}{\Delta T_{\rm{SFT}}} \quad \rm{for} \quad k = N_{\rm{bin}}/2,...,N_{\rm{bin}}-1
\end{equation}
The $l$-th time sample, $t_{l}$ spans the duration $T_{I}-\Delta T_{\rm{SFT}}/2 \leq t_{l} \leq T_{I} + \Delta T_{\rm{SFT}}/2$ where I refers to the SFT number ($I=0,1,...T_{\rm{obs}}/\Delta T_{\rm{SFT}}$), $T_{\rm{obs}}$ is the total duration of the signal, while $T_{I}$ is the central time of the SFT. While all tests in this paper make use of a Hann-window in order to reduce spectral leakage, hereafter we simplify all equations by using Equation (\ref{eq:FFTnoW}) for the SFT.

We work under the assumption that the signal $h(t)$ is quasi-monochromatic i.e., during each time interval of length $\Delta T_{\rm{SFT}}$ the signal power is, to good approximation, all contained in one single frequency bin so that:
\begin{eqnarray}
\label{eq:h(t)approx}
\nonumber h(t)\approx  h_0(T_I){\cal A}_+F_{+,I}\cos(\Phi(T_I)+2\pi f(T_I)(t-T_I))+\\
h_0(T_I){\cal A}_{\times}F_{\times,I}\sin(\Phi(T_I)+2\pi f(T_I) (t-T_I)),~~~~~~~
\end{eqnarray}
where ${\cal A}_{+},~{\cal A}_{\times}$ are amplitude factors dependent on the physical system's inclination angle $\iota$ (for on-axis GRBs, $\iota$ is the angle between the jet axis and the line of sight):
\begin{align}
{\cal A}_+&=\frac{1+\cos^2\iota}{2},\label{eq:Aplus}\\
{\cal A}_{\times}&=\cos\iota \label{eq:Across},
\end{align}
and $F_{+,I},~F_{\times,I}$ are the antenna factors that quantify a detector's sensitivity to each polarization state. We note that $\Phi(T_I)$ in Eq. (\ref{eq:h(t)approx}) may contain an unknown initial phase constant and, generally speaking, is a detector-dependent term. For simplicity, hereafter we assume that data streams from different detectors are corrected for the expected time lag in the GW signal arrival time before calculating the detection statistic. This is a reasonable assumption because our analysis focuses on searches for GWs from sources of known sky location (such as $\gamma$-ray triggered bursts). With this choice, we eliminate the dependence of $\Phi(T_I)$ on the detector and, in what follows, we do not need to introduce a detector-dependent index for the phase difference $\Delta \Phi_{IJ}$ (see Eq. (\ref{eqphasediff})). 

The raw cross-correlation statistic is defined as \cite{Dhurandhar2008}:
\begin{equation}
{\cal Y}_{IJ} = \frac{\tilde{x}^{*}_{I}[f_{k,I}]\tilde{x}_{J}[f_{k',J}]}{\Delta T^{2}_{\rm{SFT}}},
\label{eq:rawrho}
\end{equation}
 where $I$ and $J$ refer to SFT times, and $f_{k,I}$ and $f_{k',J}$ are the frequencies at which all of the signal power is concentrated during the $I$-th and $J$-th time intervals, respectively.
 The detection statistic $\rho$ can then be built as a weighted sum of the raw cross-correlation:
\begin{equation}
\rho=\sum\limits_{IJ}\limits^{N_{\rm{pairs}}}(u_{IJ}{\cal Y}_{IJ}+u^{*}_{IJ}{\cal Y}^{*}_{IJ}),
\label{eq:rhosum}
\end{equation}
where \cite{Dhurandhar2008}:
\begin{equation}
u_{IJ} = \frac{\sqrt{({\cal A}^{2}_{+}F^{2}_{+,I}+{\cal A}^{2}_{\times}F^{2}_{\times,I})({\cal A}^{2}_{+}F^{2}_{+,J}+{\cal A}^{2}_{\times}F^{2}_{\times,J})}}{\Delta T^{-2}_{\rm{SFT}}e^{-i\Delta \theta_{IJ}}S_{n}[f_{k,J}]}.
\end{equation}
In the above Equation, $\Delta \theta_{IJ}$ is defined as \citep{2016Coyne}: 
\begin{equation}
\Delta \theta_{IJ}=\pi\Delta T_{\rm{SFT}}(f_{k,I}-f_{k',J})+\Delta\Phi_{IJ},  
\end{equation}
where
\begin{equation}
\Delta\Phi_{IJ} = \Phi(T_{I}) - \Phi(T_{J})
\label{eqphasediff}
\end{equation}
(see Eq. \ref{eq:h(t)}), 
 and $S_{n}[f]$ is the single-sided power spectral density (PSD) of the detector noise which can be calculated as in Eq. 2.20 of \cite{Dhurandhar2008}
\begin{equation}
S_{n}[f_{k}] \approx \frac{2}{\Delta T_{\rm{SFT}}}\langle|\tilde{x}_{I}[f_{k}]|^{2}\rangle,
\end{equation}
where $\langle|\tilde{x}_{I}[f_{k}]|^{2}\rangle$ is the average value of the square of the transformed detector data of a given frequency (as in Eq. \ref{eq:FFTnoW} or \ref{eq:FFTW}) over a period of time $\Delta T_{\rm{SFT}}$ where the detector data may be assumed to be stationary and Gaussian.

Hereafter, we assume that the antenna  factors are constant in time throughout the duration of the GW signals here considered. However, they can vary based on detector's location and arms' orientation.
Inserting Equation (\ref{eq:rawrho}) into (\ref{eq:rhosum}) one gets:
\begin{equation}
\rho = \frac{1}{\Delta T^{2}_{\rm{SFT}}}\sum\limits_{IJ}\limits^{N_{\rm{pairs}}}u_{IJ}\tilde{x}^{*}_{I}[f_{k,I}]\tilde{x}_{J}[f_{k',J}] + u^{*}_{IJ}\tilde{x}_{I}[f_{k,I}]\tilde{x}^{*}_{J}[f_{k',J}],
\label{eq:main_rho}
\end{equation}
which shows that the distribution of $\rho$ depends on the pairs we choose to correlate. As we discuss  in what follows, with the $\rho$ statistic one can encompass various regimes, from matched-filter (fully coherent) to stochastic-like searches, with a semi-coherent approach in between. We stress that the only information needed to construct the above statistic is the signal time-frequency evolution. Thus, hereafter we refer to a model time-frequency track as a template (see Sec. \ref{Sec: Template} for more details). Generally speaking, a given time-frequency track will map onto specific physical parameters of the emitting source (see e.g. Eqs. (\ref{eq:NS-SpinDown})-(\ref{eq:freqgw}) and Figure \ref{fig:waveforms} in Section \ref{sec:testwave} for the specific case of a secularly unstable magnetar).

\subsection{Stochastic limit}
In the stochastic limit, we only correlate SFTs from different detectors (such as LIGO Hanford, LH; and LIGO Livingston, LL) at the same time (after correcting for the GW time-of-flight in case of non co-located detectors). With this choice, one minimizes computational cost and maximizes robustness against GW waveform uncertainties, at the expense of sensitivity  (when compared to e.g. the matched-filter or the semi-coherent approaches).
The number of correlated pairs in Eq. (\ref{eq:main_rho}) is $N_{\rm{pair}} = N_{\rm{SFT}}$, and we can write:
\begin{equation}
\rho = \frac{2}{\Delta T^2_{\rm{SFT}}}\sum\limits_{I}\limits^{N_{\rm{SFT}}}\Re\{u_{II}\tilde{x}^{*^{LH}}_{I}[f_{k,I}]\tilde{x}^{*^{LL}}_{I}[f_{k',I}]\}.
\label{eq:rhoSTOCH}
\end{equation}
As evident from the above equation, $\rho$ is a weighted sum of independent random variables that, under the assumption of stationary Gaussian noise, are each the product of two Gaussian variables. By the central limit theorem this sum converges to a Gaussian-distributed random variable with mean $\mu_{\rm{\rho}}$ and variance $\sigma^{2}_{\rm{\rho}}$ given by (see also Eqs. (4.17) and (4.18) in \citep{2016Coyne}):
\begin{widetext}
\begin{eqnarray}
\mu_{\rm{\rho}} = ({\cal A}^{2}_{+}F^{2}_{+,\rm{H}}+{\cal A}^{2}_{\times}F^{2}_{\times,\rm{H}})({\cal A}^{2}_{+}F^{2}_{+,\rm{L}}+{\cal A}^{2}_{\times}F^{2}_{\times,\rm{L}})\frac{\Delta T^{2}_{\rm{SFT}}}{2}\sum\limits_{I}\limits^{N_{\rm{SFT}}}\frac{h^{2}_{0}(T_{I})}{S^{LH}_{n}[f_{k,I}]S^{LL}_{n}[f_{k,I}]},
\label{eq:STOCHmu}
\end{eqnarray}

\begin{equation}
\sigma^{2}_{\rm{\rho}} = ({\cal A}^{2}_{+}F^{2}_{+,H}+{\cal A}^{2}_{\times}F^{2}_{\times,H})({\cal A}^{2}_{+}F^{2}_{+,L}+{\cal A}^{2}_{\times}F^{2}_{\times,L})\frac{\Delta T^{2}_{\rm{SFT}}}{2}\sum\limits_{I}\limits^{N_{\rm{SFT}}}\frac{1}{S^{LH}_{n}[f_{k,I}]S^{LL}_{n}[f_{k,I}]}.
\label{eq:STOCHsigma}
\end{equation}
\end{widetext}
The mean of $\rho$ is zero in the absence of a signal (assuming noise from the two detectors is uncorrelated), and has a non-zero positive value when a GW signal is present in the detectors' data.

\subsection{Matched-filter limit}
\label{sec:matched-filter}
In the matched-filter limit, we correlate all possible SFT pairs (including self-pairs), so we have $N_{\rm{pair}} = N_{\rm{SFT}}^2$, $N_{\rm{SFT}}=N_{\rm{det}}T_{\rm{obs}}/\Delta T_{\rm{SFT}}$, where $N_{\rm{det}}$ is the number of detectors from which data are taken. In this limit it can be shown that Equation (\ref{eq:rhosum}) becomes (see also Eq. (4.29) in \cite{2016Coyne}):
\begin{equation}
\rho = 2[(\sum\limits^{N_{\rm{SFT}}}_{I}\Re(\tilde{x}'_{I}[f_{k,I}]))^{2}+(\sum\limits^{N_{\rm{SFT}}}_{I}\Im(\tilde{x}'_{I}[f_{k,I}]))^{2}],
\end{equation}
where $\tilde{x}'_{I}[f_{k,I}]$ is defined as:
\begin{equation}
\tilde{x}'_{i}[f_{k,I}] = \frac{\sqrt{{\cal A}^{2}_{+}F^{2}_{+,I}+{\cal A}^{2}_{\times}F^{2}_{\times,I}}}{S_{n}[f_{k,I}]}\tilde{x}_{i}[f_{k,I}]e^{-i\theta_{i}}.
\end{equation}
For stationary Gaussian noise with zero mean, the real and imaginary parts of $\tilde{x}'_{i}$ are still Gaussian distributed, as is the case for $\tilde{x}_{i}$, and so are their sums. More specifically, in the absence of a signal, the sums of the real and imaginary parts of $\tilde{x}'_{i}$ have zero mean and variance given by:
\begin{equation}
\sigma^{2}_{\rm{\Sigma}} = \sum\limits^{N_{\rm{SFT}}}_{i}\frac{\Delta T_{\rm{SFT}}({\cal A}^{2}_{+}F^{2}_{+,I}+{\cal A}^{2}_{\times}F^{2}_{\times,I})}{4S_{n}[f_{k,I}]}=\frac{C_{\chi}}{2}.
\label{MF:Cx}
\end{equation}
Thus, $\rho$ may be re-written as the sum of the squares of two normally distributed variables, scaled by a factor $C_{\rm{\chi}}$:
\begin{equation}
\rho = C_{\rm{\chi}} \times \left[ \left(\frac{\sum\limits^{N_{\rm{SFT}}}_{I}\Re(\tilde{x}'_{I}[f_{k,I}])}{\sigma_{\rm{\Sigma}}}\right)^{2}+\left(\frac{\sum\limits^{N_{\rm{SFT}}}_{I}\Im(\tilde{x}'_{I}[f_{k,I}])}{\sigma_{\rm{\Sigma}}}\right)^{2} \right],
\label{eq:rhoMATCHED}
\end{equation}
which follows a $\chi^{2}$ distribution with two degrees of freedom, with variance and mean given by (see also Eqs. (4.36) and (4.34) in \citep{2016Coyne}):
\begin{equation}
\sigma^{2}_{\rm{\rho}} = 4C_{\rm{\chi}} = 2\sum\limits^{N_{\rm{SFT}}}_{I}\frac{\Delta T_{\rm{SFT}}({\cal A}^{2}_{+}F^{2}_{+,I}+{\cal A}^{2}_{\times}F^{2}_{\times,I})}{S_{n}[f_{k,I}]},
\label{eq:MFCx}
\end{equation}
\begin{equation}
\mu_{\rm{\rho}} = C_{\rm{\chi}}(2+\lambda),
\label{eq:MFmu}
\end{equation}
and with non-centrality parameter $\lambda$ given by (see also Eq. (4.37) in \citep{2016Coyne}): 
\begin{equation}
\lambda = \sum\limits^{N_{\rm{SFT}}}_{I}h^{2}_{0}(T_{I})\left[\frac{\Delta T_{\rm{SFT}}({\cal A}^{2}_{+}F^{2}_{+,I}+{\cal A}^{2}_{\times}F^{2}_{\times,I})}{4S_{n}[f_{k,I}]}\right].
\label{eq:MFlambda}
\end{equation}

\subsection{Semi-coherent approach}
\label{sec:semi-coh}
In the semi-coherent approach,  the total observation time $T_{\rm{obs}}$\footnote{In the case of searches for GWs associated with GRB plateaus, since we do not know the fate of the secularly unstable magnetar once it stops pumping energy into the afterglow, $T_{\rm{obs}}$ is taken to be comparable to the observed duration of the GRB X-ray plateau.} is
broken  up  into $N_{\rm{coh}}=T_{\rm{obs}}/T_{\rm{coh}}$ coherent  segments,  each  of  duration
$T_{\rm{coh}}$. The coherence time is defined as the length of time wherein the signal is expected
to maintain phase coherence (and therefore good agreement) with the model predictions. All possible SFT-pairs within each coherent time segment are cross-correlated (thus $N_{\rm{SFT}} < N_{\rm{pair}} < N^{2}_{\rm{SFT}}$, with $N_{\rm{SFT}}=N_{\rm{det}}T_{\rm{coh}}/\Delta T_{\rm{SFT}}$), and the results for each coherent time segment are then combined incoherently. 
A semi-coherent search can thus be regarded as the sum of $N_{\rm{coh}}$ matched-filter searches carried out over $N_{\rm{coh}}$ time segments each of duration $T_{\rm{coh}}$Thus, $\rho$ may be written as:
\begin{widetext}
\begin{eqnarray}
\rho = \sum\limits^{{N_{\rm{coh}}}}_M C_{\rm{\chi}_M} \times \left[ \left(\frac{\sum\limits^{N_{\rm{SFT}}/N_{\rm{coh}}}_{I}\Re(\tilde{x}'_{I \times M}[f_{k,I\times M}])}{\sigma_{{\Sigma}_M}}\right)^{2}+\left(\frac{\sum\limits^{N_{\rm{SFT}}/N_{\rm{coh}}}_{I}\Im(\tilde{x}'_{I\times M}[f_{k,I\times M}])}{\sigma_{{\Sigma}_M}}\right)^{2} \right],
\label{eq:rhosemico}
\end{eqnarray}
\end{widetext}
where $C_{{\chi}_M}$ and $\sigma_{{\Sigma}_{M}}$ are defined from Eqs. (\ref{eq:MFCx}) and (\ref{eq:MFmu}) for the duration of the M-th coherent segment only. If the PSDs of the detectors are relatively flat over the range of frequencies of interest for the searched GW signal, if their antenna factors $F_{+}$ and $F_{\times}$ are comparable (as is the case for co-located detectors with parallel arms) and slowly varying over $T_{\rm{obs}}$, then $C_{\chi_M}$ is approximately constant through each coherent segment and the resulting statistic for the cross-correlation $\rho$ is that of a $\chi^2$-distributed random variable with $2N_{\rm{coh}}$ degrees of freedom, whose variance and mean are given by (see Eqs. (4.41) and (4.42) in \cite{2016Coyne}):
\begin{equation}
\sigma^{2}_{\rm{\rho}} = 4C_{\rm{\chi}}
\end{equation}
\begin{equation}
\mu_{\rm{\rho}} = \frac{C_{\rm{\chi}}}{N_{\rm{coh}}}(2N_{\rm{coh}}+\lambda),
\label{eq:SCmu}
\end{equation}
where $\lambda$ is defined in the same way as for the matched-filter limit, Eq. (\ref{eq:MFlambda}). Note that the above equations reduce to Eqs. (\ref{eq:MFCx}) and (\ref{eq:MFmu}) for $N_{\rm{coh}}=1$, while for large $N_{\rm{coh}}$ the distribution of $\rho$ approaches a Gaussian. 

\section{Test waveforms}
\label{sec:testwave}
Throughout this paper we test CoCoA on waveforms representing GW signals that may be expected from secularly unstable, long-lived magnetars formed in GRBs (either long or short), as proposed by \cite{Corsi2009}. As discussed in Section \ref{sec:intro}, highly-magnetized NSs may be the long-lived remnants powering (via magnetic dipole losses) the  X-ray plateaus observed in GRB afterglows. Rotating NSs can also be efficient emitters of GWs if the ratio of their rotational kinetic energy to their gravitational binding energy, $\beta=T/|W|$, is in the range $0.14 < \beta <0.27$ \cite{LaiShapiro1995}. Values of $\beta$ in this interval make NSs unstable for secular bar-mode deformations whose characteristic timescales are compatible with the observed durations of GRB X-ray plateaus ($10^2-10^4$\,s). 
Under the effect of GW losses, a secularly unstable NS will follow a quasi-static
evolution along an equilibrium sequence
of tri-axial ellipsoidal figures. Adding the effect of magnetic field losses, the NS spin-down law can be written as (see Eq. 11 in \cite{Corsi2009}):
\begin{equation}
\frac{dE}{dt} = \frac{dE_{GW}}{dt} + \frac{dE_{EM}}{dt} = -\frac{B^{2}R^{6}\Omega^{4}_{\rm{eff}}}{6c^{3}}-\frac{32GI^{2}\epsilon^{2}\Omega^{6}}{5c^{5}},
\label{eq:NS-SpinDown}
\end{equation}
where $E$ is the total energy; $dE_{GW}/dt$ accounts for GW energy losses; $dE_{EM}/dt$ is the energy loss due to magnetic dipole radiation, calculated  by conserving the magnetic field flux over a sphere of radius equal to the mean stellar radius \citep[see][for more details]{Costa1997}; $B$ is the magnetic dipole field strength at the poles; $R$ is the geometric mean of the principal axes of the star; $\Omega$ is the pattern angular frequency of the ellipsoidal surface of the star; $\Omega_{\rm{eff}}$ is an effective angular frequency which includes both the ellipsoidal pattern speed and the effects of the internal fluid motions; $\epsilon=(a^2_1-a^2_2)/(a^2_1+a^2_2)$ is the ellipticity (with $a_1$ and $a_2$ as the principal axes of the ellipsoidal figure in the equatorial plane); and $I$ is the moment of inertia with respect to the star's rotation axis. The GW losses result in a quasi-periodic GW signal of frequency 
\begin{equation}
f (t) =\Omega(t)/\pi, 
\label{eq:freqgw}
\end{equation}
and amplitude given by (see Eq. (14) in \cite{Corsi2009}):
\begin{equation}
h_0(t) = \frac{4G\Omega(t)^2}{c^{4}d}I(t)\epsilon(t),
\label{eq:h(t)}
\end{equation}
where $d$ is the distance to the source. 
\label{Sec:Waveforms}
\begin{table*}
\centering
\caption{\label{tab:Wvfrm}Physical parameters, time duration, frequency range, average and maximum $\dot{f}$ through the duration of the signal, and total energy radiated in GWs and through EM dipole radiation from secularly unstable magnetars used in this study. See Section \ref{Sec:Waveforms} for more details. We use CM09long for the tests described in Section \ref{Sec:Noise}. Both CM09long and CM09short are used for the tests described in Sections \ref{sec:bank}. Bar1-6 are used for deriving the results presented in Appendix \ref{Sec: Compare} (see also  \cite{2017Postmerger}).}
\begin{tabular}{cccccccccccc}
\hline
\hline
Waveform & $\beta$ & $M$  &$R$  & $B$ & $T$  & $f_{\rm{0}}$ & $f_{\rm{f}}$ & $\left
\langle\dot{f}\right\rangle$ & $\dot{f}_{\rm{max}}$ &$E_{GW}$ & $E_{EM}$ \\
& & ($M_{\odot}$) &(km) & (Gauss) & (s) & (Hz)& (Hz) & (Hz/s) & (Hz/s) & (erg) & (erg)\\
\hline
CM09long & 0.2 & 1.4 & 20 & $10^{14}$ & 2917 & 153 & 48 & 0.05 & 0.06 & $7.6 \times 10^{50}$ & $6.0 \times 10^{50}$ \\
\hline
CM09short & 0.26 & 1.4 & 20 & $10^{14}$ & 470 & 251 & 79 & 0.60 & 3.07 & $4.1 \times 10^{51}$ & $4.6 \times 10^{49}$ \\
\hline
Bar1 & 0.2 & 2.6 &12 & $10^{13}$ & 277 & 449 & 139 & 1.21 & 7.15 & $7.9 \times 10^{51}$ & $1.7 \times 10^{48}$\\
\hline
Bar2 & 0.2 & 2.6 &14 & $10^{13}$ & 509 & 356 & 111 & 0.51 & 3.06 & $6.7 \times 10^{51}$ & $3.1 \times 10^{48}$\\
\hline
Bar3 & 0.2 & 2.6 &12 & $10^{14}$ & 237 & 449 & 139 & 1.37 & 7.19 & $7.7 \times 10^{51}$ & $1.8 \times 10^{50}$\\
\hline
Bar4 & 0.2 & 2.6 &14 & $10^{14}$ & 396 & 356 & 111 & 0.64 & 3.09 & $6.4 \times 10^{51}$ & $3.1 \times 10^{50}$\\
\hline
Bar5 & 0.2 & 2.6 &12 & $5 \times 10^{14}$ & 107 & 449 & 139 & 3.09 & 7.84 & $6.0 \times 10^{51}$ & $1.9 \times 10^{51}$\\
\hline
Bar6 & 0.2 & 2.6 &14 & $5 \times 10^{14}$ & 136 & 356 & 111 & 1.89 & 3.70 & $4.3 \times 10^{51}$& $2.5 \times 10^{51}$\\
\hline
\end{tabular}
\end{table*}
\begin{figure*}
\centering
\hbox{
\includegraphics[width=8.5cm]{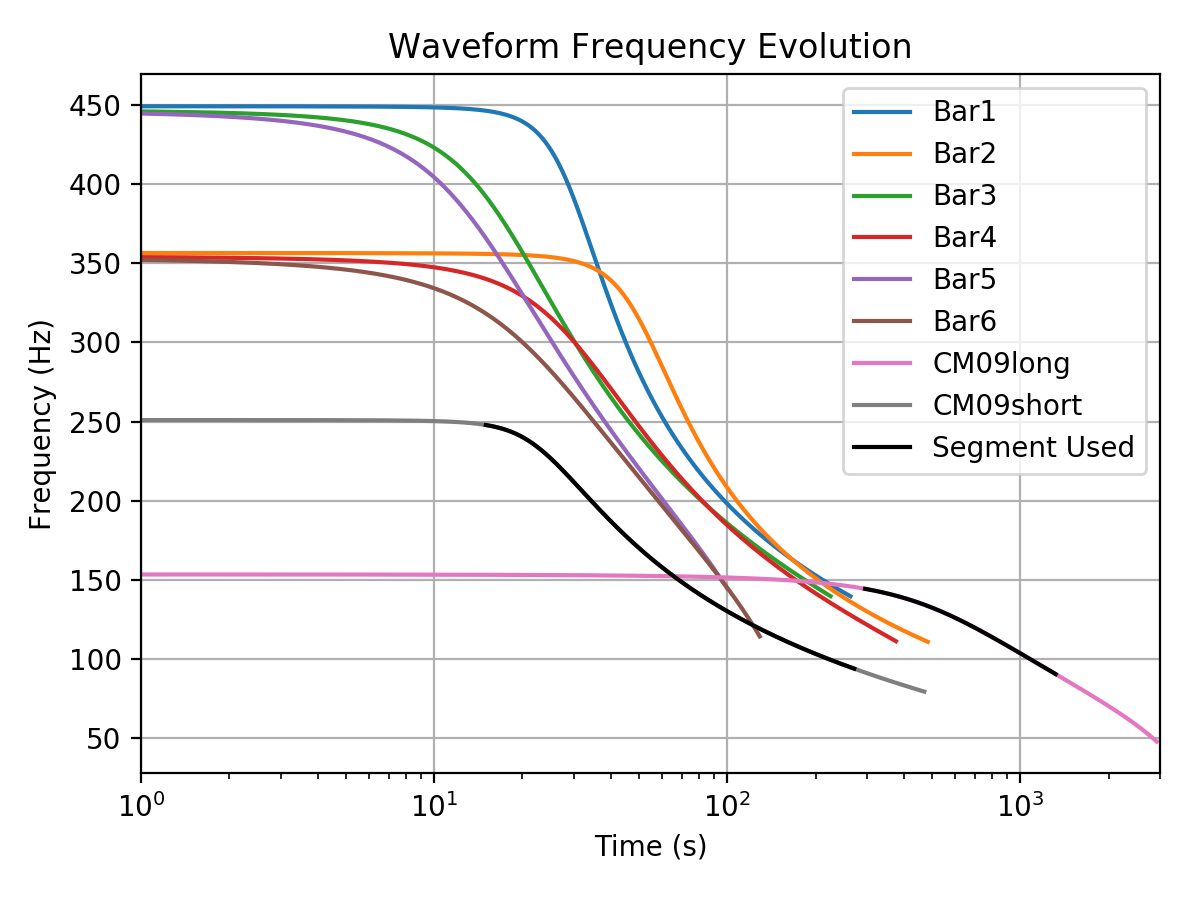}
\includegraphics[width=8.5cm]{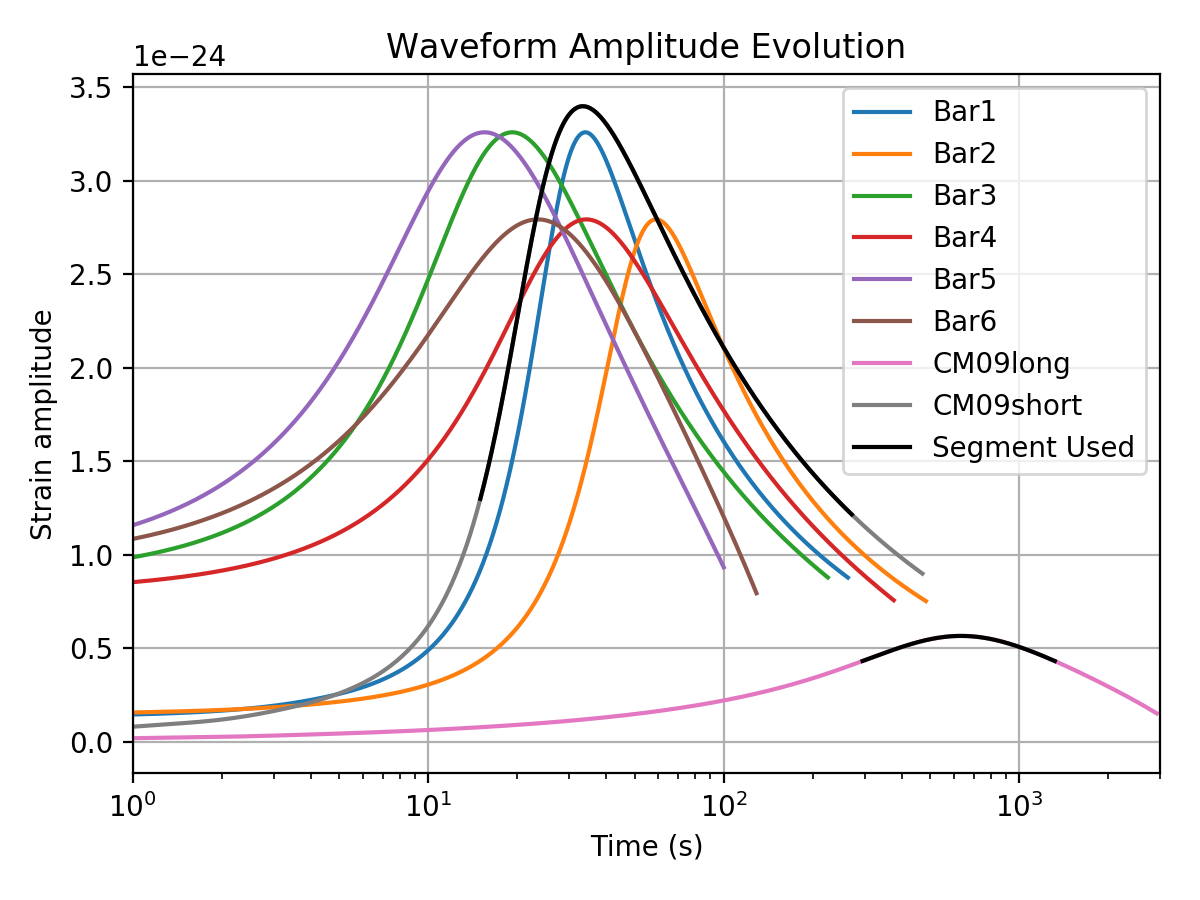}}
\caption{GW signal frequency (left) and amplitude (right) as a function of time for the waveforms used in this study (see \citet[][]{Corsi2009}, and also Section \ref{Sec:Waveforms} and Table \ref{tab:Wvfrm}). The thick black portions of the CM09short/long waveforms represent the 256/1024\,s-long segments where the sliding average of the signal strain is maximized. These portions of the CM09long/short signals are used in this study to allow for direct comparison with the results presented in \cite{2016Coyne} (see text for further discussion).\label{fig:waveforms}}
\end{figure*}

In Fig. \ref{fig:waveforms} we show the time evolution of the GW frequency $f(t)$ and strain amplitude $h_0(t)$ for signals associated with secularly unstable magnetars located at $d=100$\,Mpc, with physical parameters listed in Table \ref{tab:Wvfrm}. In this Table we also list the approximate frequency range and duration of the waveforms. Note that since in general we do not know how long a magnetar will survive before potentially collapsing to a BH, the time duration in Table \ref{tab:Wvfrm} is the time it takes for the GW luminosity to drop below 1$\%$ of its peak value (so as to enclose the bulk of the emitted GW energy, which is reported in the second to last column of this Table).

The waveform dubbed CM09long was first presented in \cite{Corsi2009}, and further used in \cite{2016Coyne} to test the performance of CoCoA on detecting such a signal when embedded in simulated white Gaussian noise. CM09short was introduced and used for similar purposes in \cite{2016Coyne}. These CM09 waveforms represent what could be a typical newly-born, rapidly-rotating NS. The initial $\beta$ for CM09long lies in the middle of the range expected for secularly unstable NSs, while the initial $\beta$ for CM09short approaches the upper bound of this range. Moreover, these waveforms span a frequency range well matched to the most sensitive portion of the LIGO PSD. In order to allow for direct comparison with the results presented in \cite{2016Coyne}, hereafter the CM09long (CM09short) waveform is further cut to consider only the 1024\,s (256\,s) where a sliding average on the signal amplitude returns the highest average strain. We use CM09long in Section \ref{Sec:Noise} to compare CoCoA performance on real LIGO data with that on simulated noise. We use both CM09 waveforms in Section \ref{sec:bank} to test the multi-trial approach of CoCoA introduced in Section \ref{Sec:Statistic}. 

Finally, in Appendix \ref{Sec: Compare} we use six waveforms first presented in the post-merger analysis of GW170817 \cite{2017Postmerger}, so as to allow for a more direct comparison of the CoCoA algorithm with other GW data analysis techniques described in \citep{2017Postmerger}. All of these waveforms assume the same NS mass of $2.6 M_{\odot}$ (see Table \ref{tab:Wvfrm}), close to the lower bound of the estimated total mass range for GW170817  ($2.73 M_{\odot}$), and to the lower bound for the total mass range  of other known binary systems ($2.57 M_{\odot}$; \cite{2017Abott}). Magnetic field values range from $10^{13}$ to $5 \times 10^{14}$\,Gauss (Table \ref{tab:Wvfrm}). Magnetic field strengths below $10^{13}$\,Gauss are unrealistic given the post-merger remnant dynamics which produce strong fields, while fields above $5\times 10^{14}$\,Gauss dominate the NS total energy loss, breaking down model assumptions (see  \cite{Corsi2009} for more details) and making the GW contribution irrelevant. NS radii of 12-14\,km are assumed to account for the fact that realistic equations of state would require quite large radii for a NS as heavy as $2.6 M_{\odot}$ \cite{2017Postmerger}. 

\section{Simulated Gaussian noise \lowercase{vs.} real noise performance of C\lowercase{o}C\lowercase{o}A}
\label{Sec:Noise}
In this section we test the performance of CoCoA on both real detector data (from LIGO sixth Science run, S6, and advanced LIGO first and second observing runs, O1 and O2) and simulated Gaussian noise with sensitivity matched to the nominal LIGO sensitivity (during S6, O1, or O2, see \cite{2018aLIGO}). We compare and contrast these results  with the analytical estimates discussed in Section \ref{Sec:II}.  To allow also for a direct comparison with \cite{2016Coyne}, all the tests described in this section use 1024\,s of the waveform CM09long (see Section \ref{Sec:Waveforms}), an SFT baseline of $\Delta T_{\rm{SFT}}=2$\,s, and for the semi-coherent approach, $N_{\rm{coh}}=4$. With these choices and for $N_{\rm{det}}=2$, we have $N_{\rm{SFT}}=512$ and thus $N_{\rm{pair}}=N_{\rm{det}}\times N_{\rm{SFT}}=2\times 512$ in the stochastic limit, $N_{\rm{pair}}=(N_{\rm{det}}\times N_{\rm{SFT}})^2=(2\times512)^2$ in the matched-filter limit, and $N_{\rm{pair}}=[(N_{\rm{det}}\times N_{\rm{SFT}})^2/N_{\rm{coh}}]=[(2\times512)^2/4]$ in the semi-coherent approach. 

\begin{figure}[ht]
\centering
\includegraphics[width=\linewidth]{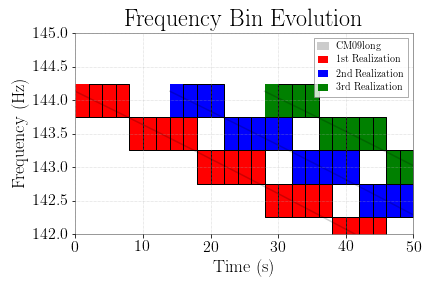}
\caption{We show how CM09long time-frequency tracks with start times 14\,s (or $7\times \Delta T_{\rm{SFT}}$) apart never overlap. Each $\Delta T_{\rm{SFT}}$ is represented with a black rectangle, and each of three non-overlapping CM09long tracks are plotted with a different color as an example. By calculating $\rho$ along each of the non-overlapping CM09long tracks that can be fitted in a given stretch of real detector data, we maximize the number of independent realizations of $\rho$ populating the statistical distributions shown in Fig. \ref{fig:S6comp}. }
\label{fig:SFT_shift}
\end{figure}

\begin{figure}
\centering
\vbox{
\includegraphics[width=8.3cm]{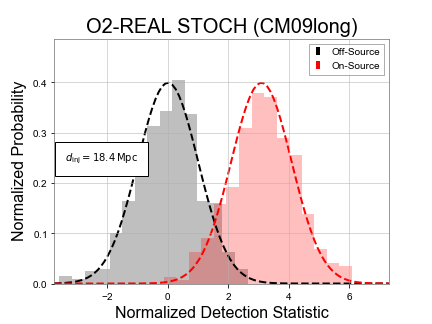}
\includegraphics[width=8.3cm]{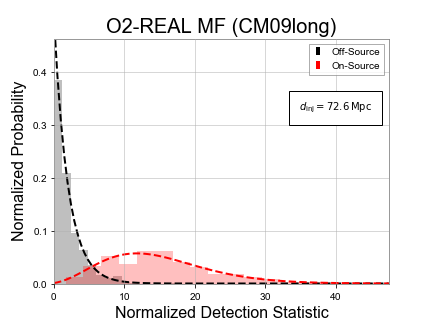}
\includegraphics[width=8.3cm]{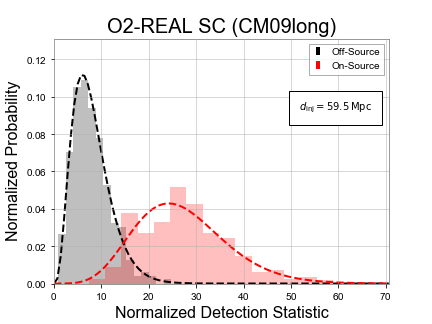}}
\caption{CoCoA tests on real O2 data from two detectors (LL and LH) in the stochastic (top), matched-filter (center), and semi-coherent  (with 4 coherent segments; bottom) limits. 
Grey histograms are background distributions normalized by the variance in Eq. (\ref{eq:STOCHsigma}) in the stochastic limit, and by $C_{\rm{\chi}}$ in Eq. (\ref{eq:MFmu}) for non-stochastic searches. Black-dashed lines are: a normalized Gaussian with zero mean (top); a central $\chi^{2}$ with 2 d.o.f. (center); a central $\chi^{2}$ with $2N_{\rm{coh}}=8$ d.o.f. (bottom). Red histograms are normalized (by $C_{\rm{\chi}}$, $C_{\rm{\chi}}/N_{\rm{coh}}$ and the variance for the matched-filter, semi-coherent and stochastic limits, respectively) distributions of real data when CM09long is injected at a distance such that $FAP$=0.1\% and a $FDP$=50\%, assuming optimal source orientation. Red lines are: a normalized Gaussian with mean as in Eq. (\ref{eq:STOCHmu}) (top); a non-central $\chi^{2}$ with 2 d.o.f. (center) and non-centrality parameter $\lambda$ as in Eq. (\ref{eq:MFlambda}); a non-central $\chi^{2}$ with $2N_{\rm{coh}}=8$ d.o.f. (bottom) and non-centrality parameter $\lambda$. \label{fig:S6comp}}
\end{figure}

The real noise tests are performed by running CoCoA on data available for public download at the LIGO Open Science Center (LOSC). Specifically, we select 6000\,s of S6 data following the GPS time 946030004, 15000\,s of O1 data following the GPS time 1132937620, and 15000\,s of O2 data following the GPS time 1186923047. These represent long segments of detector data that passed all of the basic data quality checks (cat1-3 vetoes as defined in the LOSC). We use these stretches of data to calculate the statistical distribution of $\rho$ along the 1024s-long time-frequency track of CM09long. The maximum number of independent realizations of $\rho$ obtainable from each of the S6/O1/O2 data segments is determined by the number of non-overlapping CM09long time-frequency tracks that can be fitted in such segments. This is demonstrated in Figure \ref{fig:SFT_shift}, where we show how CM09long time-frequency tracks with start times 14\,s (or $7\times \Delta T_{\rm{SFT}}$) apart never overlap. Thus, by calculating $\rho$ along each of these tracks we populate the statistical distributions shown in Fig. \ref{fig:S6comp} with more than 300 independent realizations of $\rho$ from S6 data, and almost $1000$ realizations from O1/O2 data. 

Colored Gaussian noise is generated by first simulating white Gaussian noise in the time-domain, transforming it into the frequency domain (via an SFT), scaling it by the desired PSD, and then transforming it back to the time-domain. Both real and simulated data are sampled at $f_s=4.096$\,kHz. 

We also test the performance of CoCoA when a CM09long signal is added to the data (real and simulated). To this end, for each search limit (matched-filter, stochastic, and semi-coherent) we inject CM09long at the distance where the signal at the detector has amplitude (Eq. (\ref{eq:h(t)})) such that, with a false alarm probability ($FAP$) of $0.1\%$, the false dismissal probability ($FDP$) is $50\%$ (as in \cite[][]{2016Coyne}).

\begin{figure}[ht]
\centering
\includegraphics[width=\linewidth]{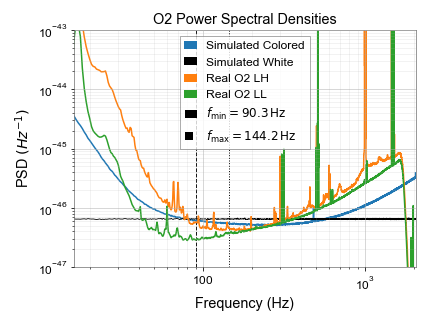}
\caption{PSDs of the LIGO O2 data used in this analysis (orange and green for LH and LL, respectively). We also plot the PSDs of the simulated colored Gaussian noise (blue) and of the simulated white Gaussian noise (black) that we use for comparison. The vertical dashed and dotted lines mark the frequency range spanned by the CM09long GW signal used in these tests (see Section \ref{Sec:Waveforms} for discussion).} 
\label{fig:PSDs}
\end{figure}

As evident from Table \ref{tab:STOCHsingle} and Figure \ref{fig:S6comp}, we find relatively good agreement (within $\approx 10$\%) of the recovered parameters of the CoCoA detection statistic on real data, simulated colored noise, and simulated white Gaussian noise (in both the absence and presence of a signal), with the analytical predictions described in Section \ref{Sec:II}.

\begin{table}[hbt!]
\caption{\label{tab:STOCHsingle} }Ratio between analytical (Section \ref{Sec:II}) and recovered values of the $\rho$ statistic for: simulated white Gaussian noise matched to LIGO S6, O1, and O2 sensitivities in the frequency range spanned by CM09long (Fig. \ref{fig:PSDs}, black); simulated colored Gaussian noise matched to S6, O1, and O2 sensitivities (Fig. \ref{fig:PSDs}, blue); real LIGO S6, O1, and O2 data (Fig. \ref{fig:PSDs}, orange and green). We note that in the matched-filter and semi-coherent limits the recovered number of d.o.f. is also consistent with the expectations of 2 and $2N_{\rm{coh}}=8$, respectively, within $15\%$.
\begin{center}
\begin{tabular}{cccc}
\hline
\hline
\multicolumn{4}{c}{\textbf{Stochastic limit}}\\
PSD & $\sigma_{\rm{\rho}}/ \sigma_{\rm{\rho},\rm rec}$ &  $\sigma_{\rm{\rho}}/ \sigma_{\rm{\rho},\rm rec}$ & $\mu_{\rm{\rho}}/ \mu_{\rm{\rho},\rm rec}$ \\
 & (noise only) & (noise+signal) &(noise+signal) \\
\hline
White S6 & 1.00 & 0.93 & 1.04\\
\hline
Colored S6 & 0.98 & 0.92 & 0.98\\
\hline
Real S6 &0.90 & 0.96 & 0.98\\
\hline
White O1 & 0.99 & 1.09 & 0.99\\
\hline
Colored O1 & 1.03 & 1.13 & 0.99 \\
\hline
Real O1 & 1.08 & 1.09 & 0.99\\
\hline
White O2 & 0.98 & 1.07 & 0.99\\
\hline
Colored O2 & 1.01 & 1.11 & 1.02 \\
\hline
Real O2 & 1.05 & 1.09 & 1.02 \\
\hline
\multicolumn{4}{c}{\textbf{Matched-filter limit}}\\
PSD & $C_{\rm{\chi}}/C_{\rm{\chi},\rm rec}$ & $C_{\rm{\chi}}/C_{\rm{\chi},\rm rec}$  & $\lambda/\lambda_{\rm rec}$ \\
 & (noise only) & (noise+signal) &(noise+signal) \\
\hline
White S6 & 0.99 & 1.13 & 0.97 \\
\hline
Colored S6 & 0.98 & 1.00 & 1.06\\
\hline
Real S6 & 0.97 & 0.95 & 1.02\\
\hline
White O1 & 1.04 & 1.01 & 1.02\\
\hline
Colored O1 & 0.95 & 1.06 & 0.98 \\
\hline
Real O1 & 1.11 & 0.96 & 0.94\\
\hline
White O2 & 0.91 & 0.93 & 0.99\\
\hline
Colored O2 & 1.00 & 1.00 & 1.02 \\
\hline
Real O2 & 1.04 & 0.96 & 1.02 \\
\hline
\multicolumn{4}{c}{\textbf{Semi-coherent approach} ($N_{\rm coh}=4$)}\\
PSD & $C_{\rm{\chi}}/C_{\rm{\chi},\rm rec}$ & $C_{\rm{\chi}}/C_{\rm{\chi},\rm rec}$  & $\lambda/\lambda_{\rm rec}$ \\
 & (noise only) & (noise+signal) &(noise+signal) \\
\hline
White S6 & 1.00 & 1.12 & 0.91\\
\hline
Colored S6 & 1.12 & 0.97 & 1.09\\
\hline
Real S6 & 1.02 & 0.84 & 1.10\\
\hline
White O1 & 1.05 & 1.02 & 1.03\\
\hline
Colored O1 & 1.09 & 1.02 & 1.03 \\
\hline
Real O1 & 1.04 & 1.09 & 0.97\\
\hline
White O2 & 0.99 & 0.99 & 0.94\\
\hline
Colored O2 & 1.00 & 1.06 & 1.01 \\
\hline
Real O2 & 1.02 & 0.90 & 1.03 \\
\hline
\end{tabular}
\end{center}
\end{table}

\section{Multi-trial search for GRB remnants}
\label{sec:multi}
In a realistic search for GWs from GRB remnants, the large uncertainties that affect the post-merger / post-explosion physics need to be taken into account. Even though CoCoA allows tuning of sensitivity/robustness so that some degree of uncertainty can be tolerated on the expected time-frequency track of the GW signal (see Section \ref{Sec:II}), larger departures from such a track would cause the search to fail. In this Section we address the need for a large parameter space exploration, give an order-of-magnitude estimate for the implied computational cost of a search spanning such space, and describe the practical implementation of a multi-trial detection statistic for CoCoA.

\label{Sec: Template}
\subsection{Remnant properties and timing uncertainties}
\label{sec:timingunc}
For the specific case of GWs from bar-mode instabilities of rotating magnetars discussed in Section \ref{Sec:Waveforms}, a realistic search with CoCoA should be performed over a template bank spanning the possible range of parameters ($\beta$, $M$, $R$, $B$, $t_{\rm on}$), where $t_{\rm on}$ accounts for the uncertainty on the  onset time of the secular bar-mode instability, something not considered in \cite{2016Coyne}. 

In Coyne et al.  \cite{2016Coyne} we have shown that, for searches based on CM09long, the maximum errors one could tolerate on the assumed magnetar properties are of the order of
$\Delta M\approx 5\times10^{-3}$\,M$_{\odot}$, $\Delta B\approx 10^{12}$\,Gauss, $\Delta R\approx 2\times10^{-2}$\,km. With these errors, the sensitivity of a CoCoA semi-coherent search with optimized $T_{\rm{coh}}$ approaches that of a stochastic search on a perfectly matching template\footnote{The last is also comparable to the maximum sensitivity of more robust and less computationally expensive algorithms that don't rely on any prior knowledge of the signal time-frequency evolution (e.g. STAMP; see \cite{2017Postmerger}).}.

For GRBs observed on-axis and forming a long-lived, secularly unstable magnetar, the expected X-ray plateau duration and luminosity depend on the initial values of $\beta$, $B$, and $R$, which can thus be constrained to some specific ranges by comparison with the observations \cite{2019Tang}. Moreover, as demonstrated in the case of GW170817 \cite{2017Postmerger}, for short GRBs some constraints on the remnant mass $M$ can be derived from the analysis of the pre-merger signal itself. The optimal case, of course, would be that of a short GRB with an observed X-ray plateau for which an in-spiral signal is also detected. In this case, joint electromagnetic and GW observations would enable us to set some constraints on all relevant parameters.

Regarding the uncertainty on $t_{\rm{on}}$, for long GRBs formed from collapsing massive stars we can reasonably assume that the delay between the collapse (and formation of the remnant) and that of the GRB trigger itself is of order 120\,s \cite{2017GRBSearch,2012GRBSearch}.
Thus, $t_{\rm GRB}-120$\,s$\,\lesssim t_{\rm collapse}\lesssim t_{\rm{on}}\lesssim t_{\rm{GRB}}$ (where $t_{\rm{GRB}}$ is the GRB trigger time in $\gamma$-rays). In the case of short GRBs from merger of compact objects, the delay between the merger and the GRB trigger time is expected to be of the order of a few seconds, thus we assume $t_{\rm GRB}-6$\,s$\,\lesssim t_{\rm merger}\lesssim t_{\rm{on}}\lesssim t_{\rm{GRB}}$ \cite{2017GRBSearch,2012GRBSearch}.
The timing uncertainty on $t_{\rm on}$ may be further reduced when  $t_{\rm{collapse}}$ or $t_{\rm{merger}}$ are more distinctly known through the detection of GWs produced by the merger/collapse. This was the case for GW170817, in which $t_{\rm GRB}-t_{\rm merger}=1.74\pm 0.05$\,s \cite{2017MultiMessenger}.

Motivated by the above considerations, in this analysis we vary $t_{\rm{on}}$ between $t_{\rm{GRB}}-t_{\rm{unc}}$ (where $t_{\rm{unc}}$ is 120\,s for long GRBs and $2-6$\,s for short GRBs) and  $t_{\rm{GRB}}$, in steps of $\Delta t_{\rm{on}}=N_{\rm{on}}\Delta T_{\rm{SFT}}$. 

\subsection{Computational cost: Order-of-magnitude estimate}
\label{sec:compcost}
As an order of magnitude estimate of the computational cost for a multi-trial CoCoA search that accounts for the uncertainties described in the previous Section, let us consider a post-merger search similar to that performed by \citep{2017Postmerger} for GW170817. The last assumed fixed values of $\beta$ and $M$, and large uncertainties in $B$ and $R$ (see also Bar1-Bar6 in Table \ref{tab:Wvfrm}). With a parameter space resolution of $\Delta B\approx 10^{12}$\,G and $\Delta R\approx 2\times10^{-2}$\,km for the magnetic field and NS radius, respectively, ranges of $B=10^{13}-5\times 10^{14}$\,G and $R=12-14$\,km, (see Section \ref{Sec:Waveforms} for discussion of this range) could be spanned with a total of $\sim 5 \times 10^4$ templates. With $t_{\rm{unc}} = 2$\,s and $\Delta t_{\rm{on}} = 0.25$\,s we take ~10 trials to cover the full timing uncertainty, making the total number of templates become $\sim 5 \times 10^5$ (see Section \ref{sec:multi-trial-ton-test}).

A search with $FAP$ of $1\%$ requires running on order 2500 background realizations per template. This number of realizations ensures that the $\rho$ probability distribution above the $FAP$ threshold is populated with 25 events, thus resulting in an error of $\approx 20\%$ for the corresponding detection efficiency. 

A CoCoA search on a single time-frequency track (as described in Section \ref{Sec:II}) with $FAP$ of $1\%$ and 2500 background realizations
is estimated to require  $\sim 1.5$ core-hours or ``Standard Units'' (SUs\footnote{An ``SU'' is an XSEDE Service Unit on Stampede, equal to 1 CPU core-hour on a 2.7 GHz E5-2680 Intel Xeon (Sandy Bridge) processor. E.g., a 1 hour allocation on a 8-core Stampede CPU would consume 8\,SUs.}). So a GRB search at $1\%$ $FAP$ on a template bank with $\sim 5 \times 10^5$ time-frequency tracks would require $\sim 0.75$\,MSUs. Assuming $\lesssim 2-3$ potentially nearby GRBs with X-ray plateaus and good LIGO Hanford/Livingston data in a 1\,yr run, we estimate a full-run multi-trial GRB search to require $(1.5-2.25)$\,MSUs, which is similar to the computational cost of other LIGO searches (e.g.,  \cite{2019Caride}).

For a two-detector CoCoA search with a template bank similar to the GW170817 post-merger analysis described here, constructing 2500 independent background realizations per template requires $\gtrsim 10.5$\,d of coincident background data. This is comparable to e.g. what was used in \cite{2017Postmerger}, where 5.6\,d of background data were derived  from non-continuous stretches of LL and LH coincident data from 2017 August 13-21 UT. We estimate that the SFTs of a $\approx 10.5$\,d-long stretch of data will consume $\sim$ 80\,GB of disk space per detector.

\subsection{Multi-trial detection statistic}
\label{Sec:Statistic}
When uncertainties on the signal properties are large and searching over multiple time-frequency tracks (template bank) becomes necessary, the detection statistic of CoCoA needs to be modified to account for the larger number of trials. Hereafter, a CoCoA search on a single template in a bank (see Section \ref{Sec:II}) will be referred to as a single trial.

To cover a given template bank, one performs a total of $N_{\rm{trial}}$  searches, each returning a certain value of the single-trial $\rho$ statistic defined as in Section \ref{Sec:II}. In general, the probability distribution of $\rho$ changes across the template bank because it depends on the properties of the time-frequency tracks that constitute the bank itself. It is thus convenient to introduce a normalized $\rho$ statistic, which in the stochastic limit we define as
\begin{equation}
\tilde{\rho}_m = \frac{\rho_{ m}}{\sigma_{{\rho,m}}}\label{eqnorm:stoch},
\end{equation}
where $\rho_m$ is a Gaussian random variable calculated along the m-th template as in Eq. (\ref{eq:rhoSTOCH}), with mean and standard deviation $\mu_{\rho,m}$ and $\sigma_{\rho,m}$ given by Eqs. (\ref{eq:STOCHmu}) and (\ref{eq:STOCHsigma}), respectively. In the matched filter limit, we define the normalized statistic as:
\begin{equation}
\tilde{\rho}_m = \frac{\rho_{ m}}{C_{{\chi},m}}\label{eqnorm:mf},
\end{equation}
where $C_{\chi,m}$ and $\rho_m$ are calculated along the m-th template as in Eqs. (\ref{MF:Cx}) and (\ref{eq:rhoMATCHED}), respectively, with $\rho_m$ a random variable distributed as a $\chi^2$ with two degrees of freedom and with non-centrality parameter $\lambda_m$ given by Eq. (\ref{eq:MFlambda}). Finally, in the semi-coherent approach we set:
\begin{equation}
\tilde{\rho}_m = \frac{N_{\rm{coh}}\rho_m}{C_{{\chi,m}}}\label{eqnorm:sc},
\end{equation}
where $C_{\chi,m}$ and $\rho_m$ are calculated along the $m$-th template as in Eqs. (\ref{MF:Cx}) and Eq. (\ref{eq:rhosemico}), respectively, with $\rho_m$ a random variable distributed as a $\chi^2$ with $2N_{coh}$ degrees of freedom, and non-centrality parameter $\lambda_m$ given by Eq. (\ref{eq:MFlambda}). In the above definition, we also assume that the same $N_{\rm coh}$ is adopted (and optimized) across trials (see Section \ref{sec:bank}). 

With the above normalization, we define the maximum $\rho$ statistic as $\tilde{\rho}_{\rm{max}}$=max$(\tilde{\rho}_m)$ for $m=0, ...,N_{\rm trial }-1$, which we use to identify the most statistically promising detection candidates. Generally speaking, in the presence of a signal, we expect the template that is most similar to the signal to return the maximum value of the $\tilde{\rho}_m$ statistic. For a given choice of $FAP$, we thus set the corresponding detection threshold as:
\begin{equation}
\int_{\tilde{\rho}_{\rm{max}} \geq \tilde{\rho}_{\rm{th}}}P(\tilde{\rho}_{\rm{max}})d\tilde{\rho}_{\rm{max}}=FAP(\tilde{\rho}_{\rm th}),
\label{eq:FAP}
\end{equation}
where $P(\tilde{\rho}_{\rm{max}})$ is the probability that any of the templates in the bank (trial) returns the largest value $\tilde{\rho}_{\rm{max}}$. For completely independent trials, this probability reads:
\begin{equation}
P(\tilde{\rho}_{\rm{max}}) = \sum\limits^{N_{\rm{trial}}}_{\rm{n=1}}\left[p_{n}(\tilde{\rho}_{\rm{max}})\prod\limits^{N_{\rm{trial}}-1}_{m\neq n}\int\limits_{\tilde{\rho}_m
\leq\tilde{\rho}_{\rm{max}}}p_{{m}}(\tilde{\rho}_{\ m})d\tilde{\rho}_{{m}}\right],
\label{eq:stochcombinedprob}
\end{equation}
where $p_m$ is the probability distribution of $\tilde{\rho}_m$.

In the absence of a signal, the probability distribution of the normalized $\rho$ statistic is the same for all trials, thus $p_m(\tilde{\rho}_m)=p(\tilde{\rho}_m)$ for all $m$ and above Equation simplifies to \cite{2010FirstSearch}:
\begin{equation}
P(\tilde{\rho}_{\rm{max}}) = N_{\rm{trial}} \times p(\tilde{\rho}_{\rm{max}})\left(\int\limits_{~\tilde{\rho}_m\leq \tilde{\rho}_{\rm{max}}}p(\tilde{\rho}_m)d\tilde{\rho}_m\right)^{N_{\rm{trial}}-1}
\label{eq:5.6}
\end{equation}
Integrating both sides of the above Equation, and considering Eq. (\ref{eq:FAP}), we get:
\begin{eqnarray}
\nonumber \frac{FAP(\tilde{\rho}_{\rm th})}{N_{\rm trial}}= ~~~~~~~~~~~~~~~~~~~\\\nonumber\int_{\tilde{\rho}_{\rm max}\geq \tilde{\rho}_{\rm th}}p(\tilde{\rho}_{\rm max})\left(1-\int\limits_{~\tilde{\rho}_m\ge \tilde{\rho}_{\rm{max}}}p(\tilde{\rho}_m)d\tilde{\rho}_m\right)^{N_{\rm trial}-1}\,d\tilde{\rho}_{\rm max}.\\
\label{eq:combinedprobsimplified}
\end{eqnarray}
If the FAP is small, then $\int\limits_{~\tilde{\rho}_m\ge \tilde{\rho}_{\rm{max}}}p(\tilde{\rho}_m)d\tilde{\rho}_m \ll 1$ for $\tilde{\rho}_{\rm max} \geq \tilde{\rho}_{\rm th}$ and we can approximate  Eq. (\ref{eq:combinedprobsimplified}) as: 
\begin{equation}
\frac{FAP(\tilde{\rho}_{\rm th})}{N_{\rm{trial}}}\approx \int\limits_{\tilde{\rho}_{\rm max}\geq \tilde{\rho}_{\rm{th}}}p(\tilde{\rho}_{\rm max})d\tilde{\rho}_{\rm max}=FAP_{\rm single\,trial}(\tilde{\rho}_{\rm th}).\label{eq:combined$FAP$calculated}\end{equation} 

The above approximation is useful as it shows that the $FAP$ threshold of a multi-trial search can be estimated analytically from the $FAP$ threshold of a single trial search. We finally note that for searches where the individual trials are not fully independent, the probability of a given $\tilde{\rho}_{\rm{max}}$ is generally lower than what is predicted in Equation (\ref{eq:stochcombinedprob}), so one can define an effective number of trials $N_{\rm{eff,trials}}\lesssim N_{\rm{trial}}$. In the case of GRB magnetars, templates are fully independent only when their time-frequency tracks as determined by ($\beta$, $M$, $R$, $B$, $t_{\rm{on}}$) do not intersect.

\section{C\lowercase{o}C\lowercase{o}A multi-trial search tests}
\label{sec:bank}
As discussed in Section \ref{sec:multi}, in a post-GRB search for GWs from secularly unstable magnetars, it is necessary to build a multi-trial search that accounts not only for the uncertainties on the magnetar properties ($\beta$, $M$, $R$, $B$), but also for the uncertainty that affects the timing between the GRB trigger time as established by $\gamma$-ray observations, and the onset of the bar-mode instability. Hereafter, we present the results of tests aimed at verifying the agreement between the analytical expectations for the CoCoA multi-trial statistic described in Section \ref{sec:multi} and the actual code performance on simulated data, as well as demonstrating the sensitivity of a CoCoA search. Our tests proceed as follows:
\begin{enumerate}
\item We simulate colored Gaussian noise with PSD matching that of LIGO O2, sampled at $f_s=4.096$\,kHz. We assume two detectors with identical PSDs, use $\Delta T_{\rm{SFT}}=0.25$\,s, and set $FAP$=1\% for determining our detection threshold (see Eq. (\ref{eq:FAP})).

\item We simulate a region of data extending between $t_{\rm{GRB}}-t_{\rm{unc}}$ and $t_{\rm{GRB}}+t_{\rm{waveform}}$, where $t_{\rm{GRB}}$ is an arbitrary GRB trigger time, $t_{\rm{unc}}$ is the timing uncertainty between the collapse/merger and the GRB trigger time and $t_{\rm{waveform}}$ is the duration of the waveform being searched for. We take two values for $t_{\rm{unc}}$, 120\,s to simulate a standard long GRB and 2\,s to simulate an event similar to GW170817 (see Section \ref{sec:compcost}).
\item We assume a known GRB sky location and set $F_{+}=-0.092$ and $F_{\times}=-0.91$ for LIGO Hanford, and $F_{+}=0.26$ and $F_{\times}=0.79$ for LIGO Livingston (comparable to those of GW170817 \cite{2017Abott,2017Postmerger}).
\item When constructing a template bank for the search  we vary $t_{\rm{on}}$ in the range $[(t_{\rm{GRB}}-t_{\rm{unc}}), t_{\rm{GRB}}]$ (see Section \ref{sec:timingunc}) for each choice of $(\beta, M, R, B)$. We sample this range in steps $\Delta t_{\rm on}$ that are multiples of $\Delta T_{\rm{SFT}}$ i.e., $ \Delta t_{\rm{on}}=N_{\rm{on}} \times \Delta T_{\rm{SFT}}$. The choice of $\Delta t_{\rm{on}}$ is made with computational cost in mind given that the smaller $\Delta t_{\rm{on}}$, the larger the number $t_{\rm{unc}}$/$\Delta t_{\rm{on}}+1$ of templates required to account for the timing uncertainty. 

\item To estimate our detection efficiency, we inject signals in the simulated O2 data   assuming we are aligned with the GRB jet axis (i.e., $\iota=0$ in Eqs. (\ref{eq:Aplus})-(\ref{eq:Across})), as expected for GRBs with X-ray plateaus. The injection time $t_{\rm{inj}}$ is set to always fall exactly in between the onset times of two randomly chosen, temporally adjacent templates in the bank, i.e.  $t_{\rm{inj}}-t_{\rm{on,n}}=t_{\rm{on,n+1}}-t_{\rm{inj}}=\Delta t_{\rm{on}}/2=N_{\rm{on}} \times \Delta T_{\rm{SFT}}/2$.  With this choice, we maximize the temporal mismatch between the injected signal and the closest template in the bank, thus obtaining a conservative estimate of CoCoA's detection efficiency. 
\item Similarly to what is done in \cite{2016Coyne}, for each waveform we calculate the detection efficiency as a function of luminosity distance, and derive a distance horizon by requiring a false dismissal probability $FDP=50\%$ (see Figure \ref{fig:sygmoid}).
\end{enumerate}

\subsection{Timing uncertainties}
\label{sec:multi-trial-ton-test}
In order to first isolate the effects of timing uncertainties only, here we carry out a multi-trial CoCoA search where the signal we search for is assumed to be produced by a magnetar with exactly known parameters ($\beta$, $M$, $R$, $B$), but with unknown onset time $t_{\rm{on}}$. We thus define a template bank composed of CM09long-like/CM09short-like waveforms (see Section \ref{Sec:Waveforms}) whose onset time is varied as described in the previous Section. 

\subsubsection{Background statistic}
\label{Sec:CombinedBKG}
\begin{figure*}
\centering
\hbox{
\includegraphics[width=8.5cm]{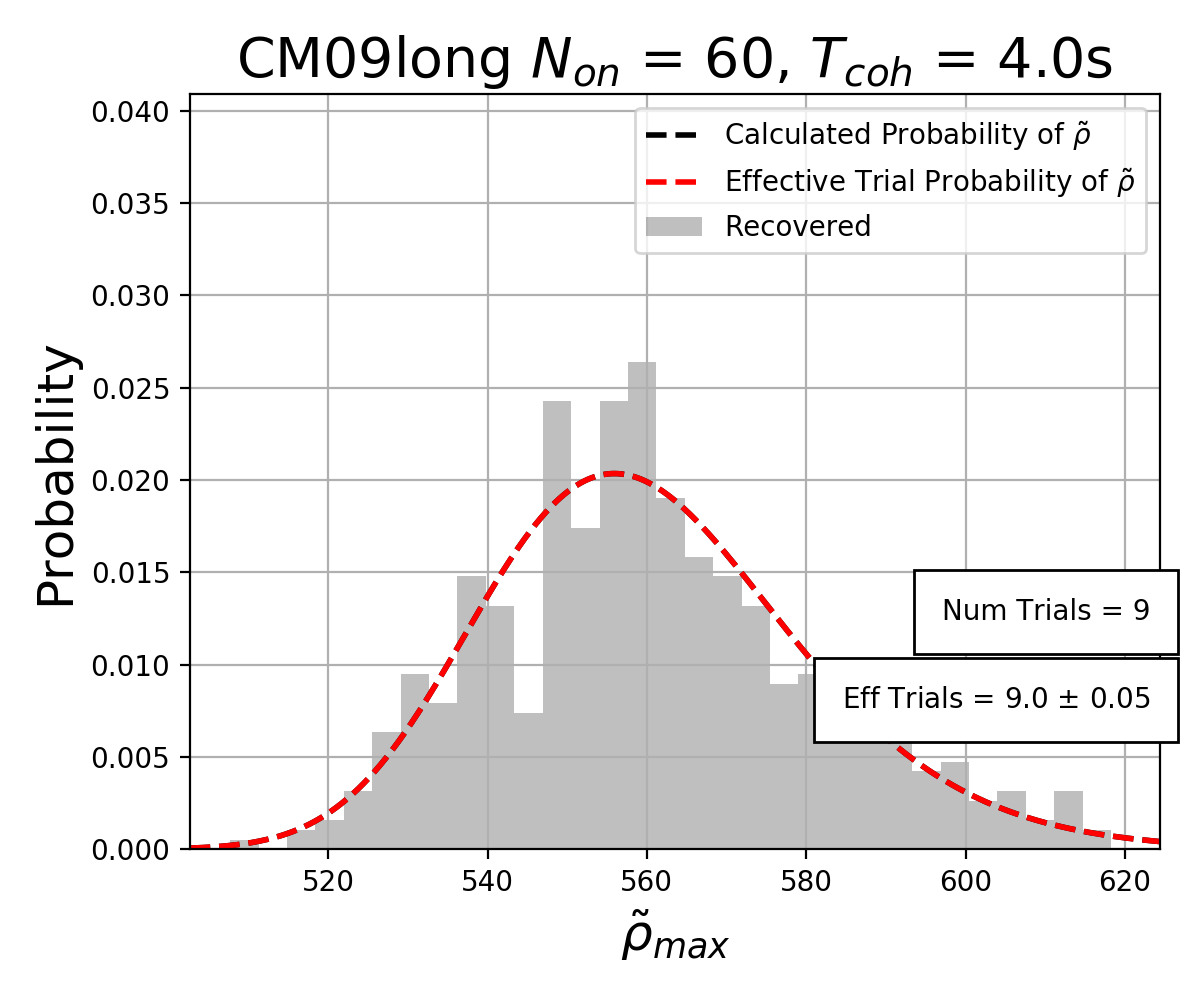}
\includegraphics[width=8.5cm]{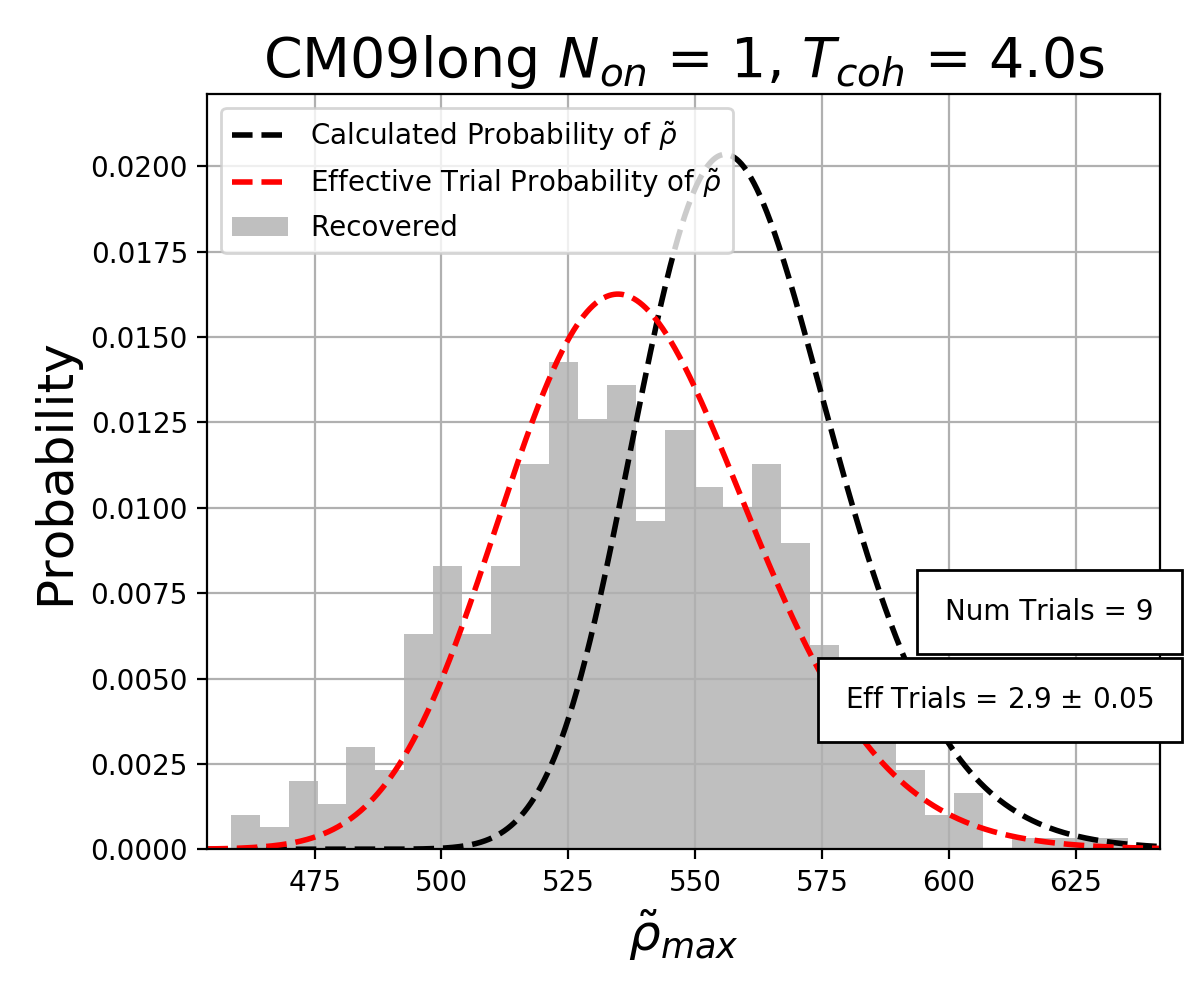}}
\hbox{
\includegraphics[width=8.5cm]{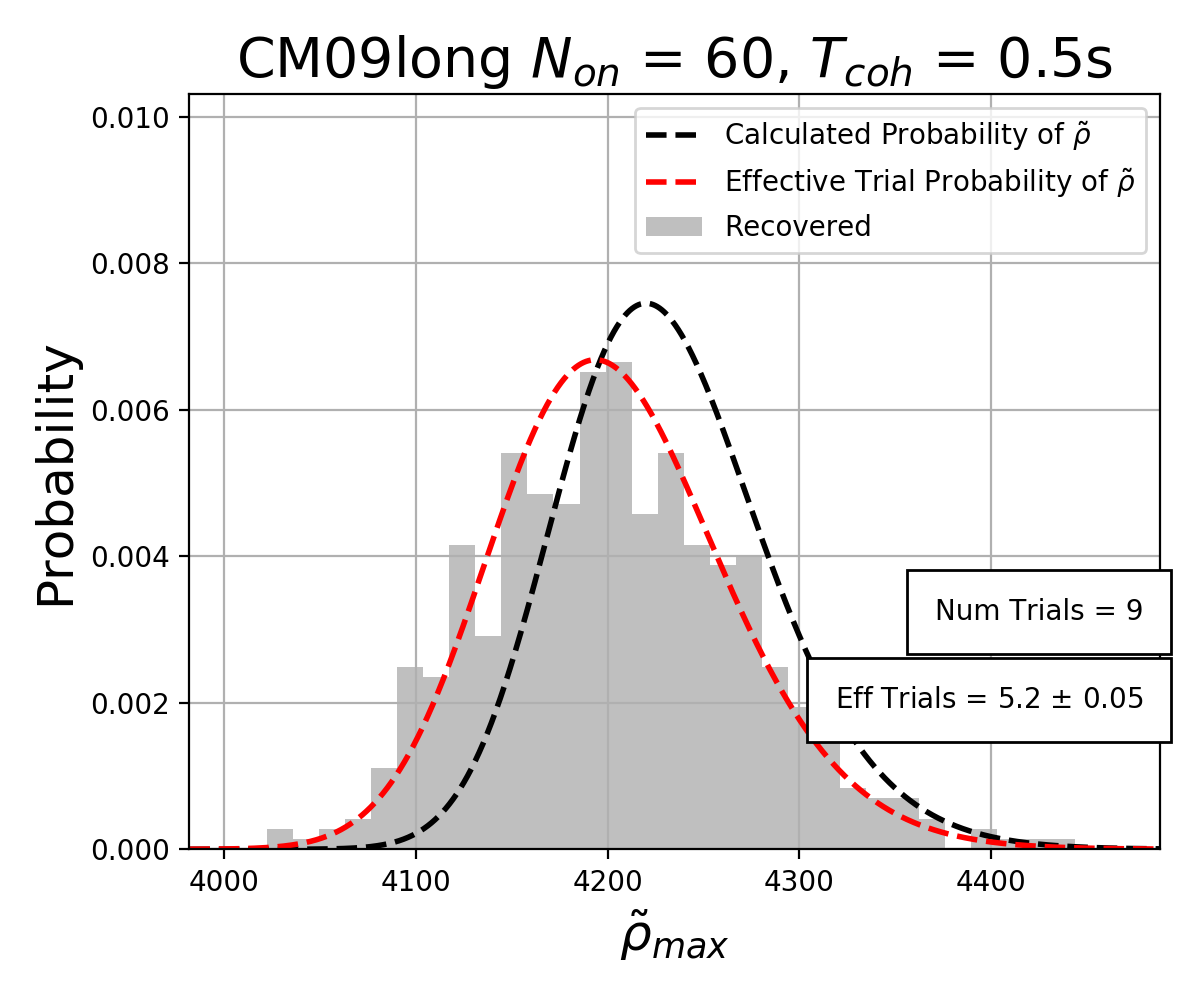}
\includegraphics[width=8.5cm]{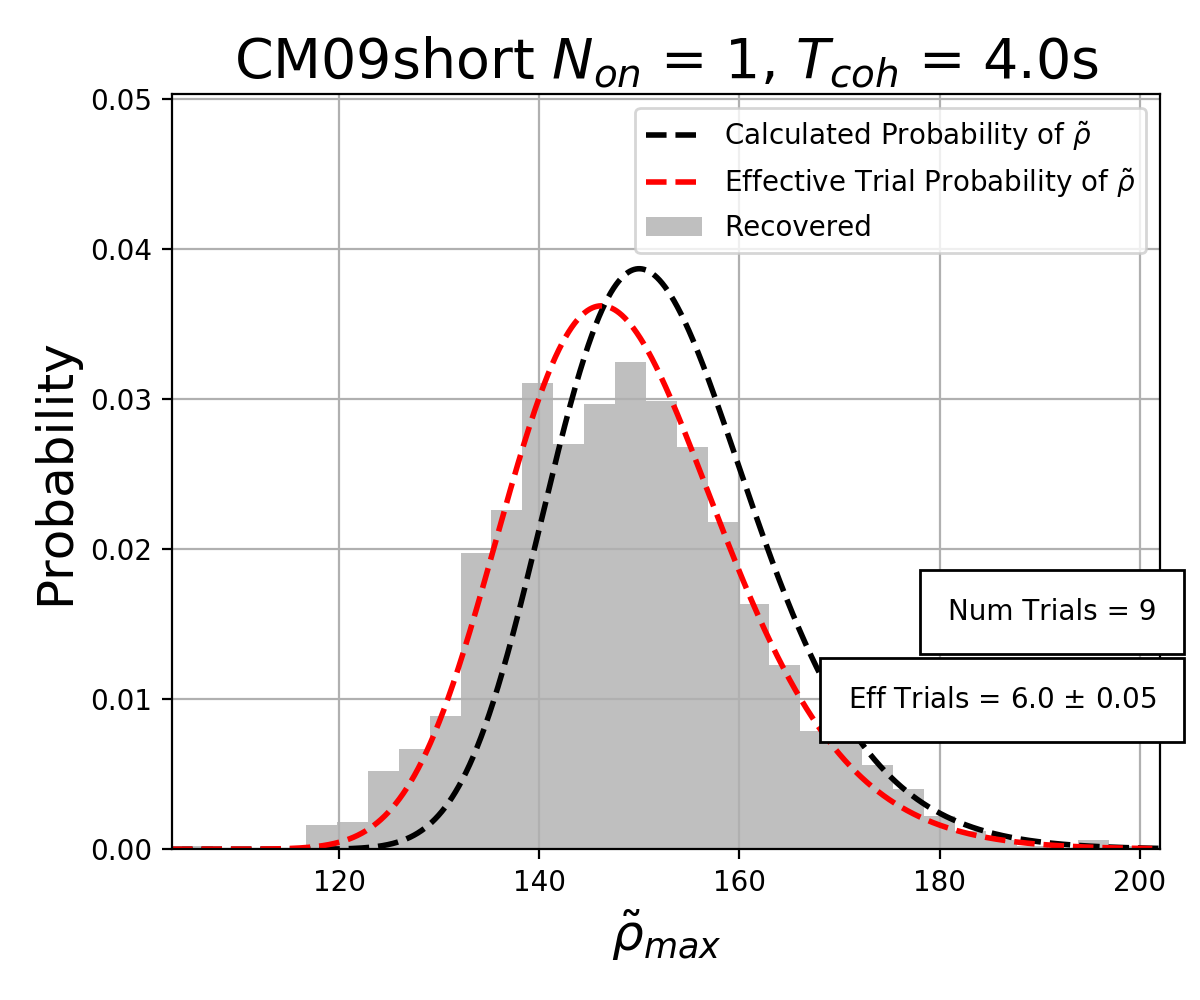}}
\caption{\label{fig:Background}Effect of timing uncertainties on a multi-trial CoCoA search: results for the background statistic. We use simulated LIGO data with (colored) O2 PSD and set $\Delta T_{\rm{SFT}}=0.25$\,s. In order to keep the number of trials fixed, for searches with  $t_{\rm{unc}}=120$\,s we take $N_{\rm{on}} = 60$, and for searches with  $t_{\rm{unc}}=2$\,s we take $N_{\rm{on}} = 1$. We compare recovered results in the absence of a signal (grey histograms) to the analytical expectations derived in Section \ref{Sec:Statistic} (black lines). To match the recovered results, we define an effective number of trials for Eq. \ref{eq:5.6} that accounts for dependencies between trials (red lines). In the first row we search for CM09long, only varying the start time of the waveform for each trial, taking identical values of $T_{\rm{coh}} = 4$\,s, but using different values of $ N_{\rm{on}}$. In the first column we search for the same waveform and use identical values of $N_{\rm{on}} = 60$, but show different values of $T_{\rm{coh}}$. Lastly in the second column we search with identical values of $N_{\rm{on}} = 1$ and $T_{\rm{coh}} = 4$\,s, but search for different waveforms (CM09long/CM09short).}
\end{figure*}
As shown in Figure \ref{fig:Background}, in the absence of a signal, the recovered multi-trial background statistic (grey histogram) for various choices of $T_{\rm{coh}}$ and $N_{\rm{on}}$ can show deviations from the analytical expectations described in Section \ref{Sec:Statistic} (black dashed line; see also Eq. (\ref{eq:5.6})). Those expectations assumed that trials are completely independent (see Eq. (\ref{eq:5.6})), which is not always the case. Indeed, varying the onset times of otherwise identical time-frequency tracks can introduce dependencies between trials, which in turn imply that the recovered background distribution is equivalent to a predicted background distribution with an effective number of trials that is lower than the one obtained assuming that all templates in the bank are independent. 

Dependencies among templates become more important for smaller values of $N_{\rm{on}}$, as evident by comparing the top-left and top-right panels of Figure \ref{fig:Background}. The recovered probability distribution (grey histogram) agrees well with the predictions discussed in Section \ref{Sec:Statistic} (black-dashed line) for large $N_{\rm{on}}$ (top-left panel). However, for smaller $N_{\rm{on}}$ in an otherwise identical search (top-right), the recovered results deviate from the expected ones. 

To describe the actual recovered background statistic for a non-fully independent template bank, we thus introduce an effective number of trials determined as described in Appendix \ref{Sec: EffTrials}. The red-dashed lines in Figure \ref{fig:Background} show that this effective background distribution agrees well with the recovered one (histogram; note that in the top-left panel the black- and red-dashed lines overlap completely).

Other factors affecting the effective number of trials include the rate at which the considered waveform evolves. For example, in a search for the faster evolving waveform CM09short, the time-frequency tracks of different trials in the bank are less likely to have significant overlaps (and thus related trial dependencies) even for small values of $N_{\rm{on}}$ (bottom-right panel in Figure \ref{fig:Background}).  Finally, smaller values of $T_{\rm{coh}}$ result in a larger degree of statistical dependence between templates (compare the top-left and bottom-left panels in Figure \ref{fig:Background}). Indeed, for a given number of overlapping time-frequency bins between two templates in a bank, the smaller the coherence time, the larger the fraction of dependent pairs (i.e. pairs generated from cross-correlation products containing time-frequency bins in the overlapping portion of the templates time-frequency track)  to the total number of pairs entering in the computation of $\rho$ along each template. Incidentally we note that, conceptually, this effect is similar to what is behind the larger robustness of semi-coherent searches with smaller coherence timescales: if only a few time-frequency bins overlap between the injected signal and the closest template in a bank, smaller coherence times imply that cross-correlation products from these few overlapping bins have a larger relative weight in the computation of $\rho$ along the template.

\subsubsection{Detection efficiency and search sensitivity}
\begin{figure*}
\centering
\textbf{$t_{\rm{unc}}$ = 120\,s}\par\medskip
\hbox{
\includegraphics[width=8.5cm]{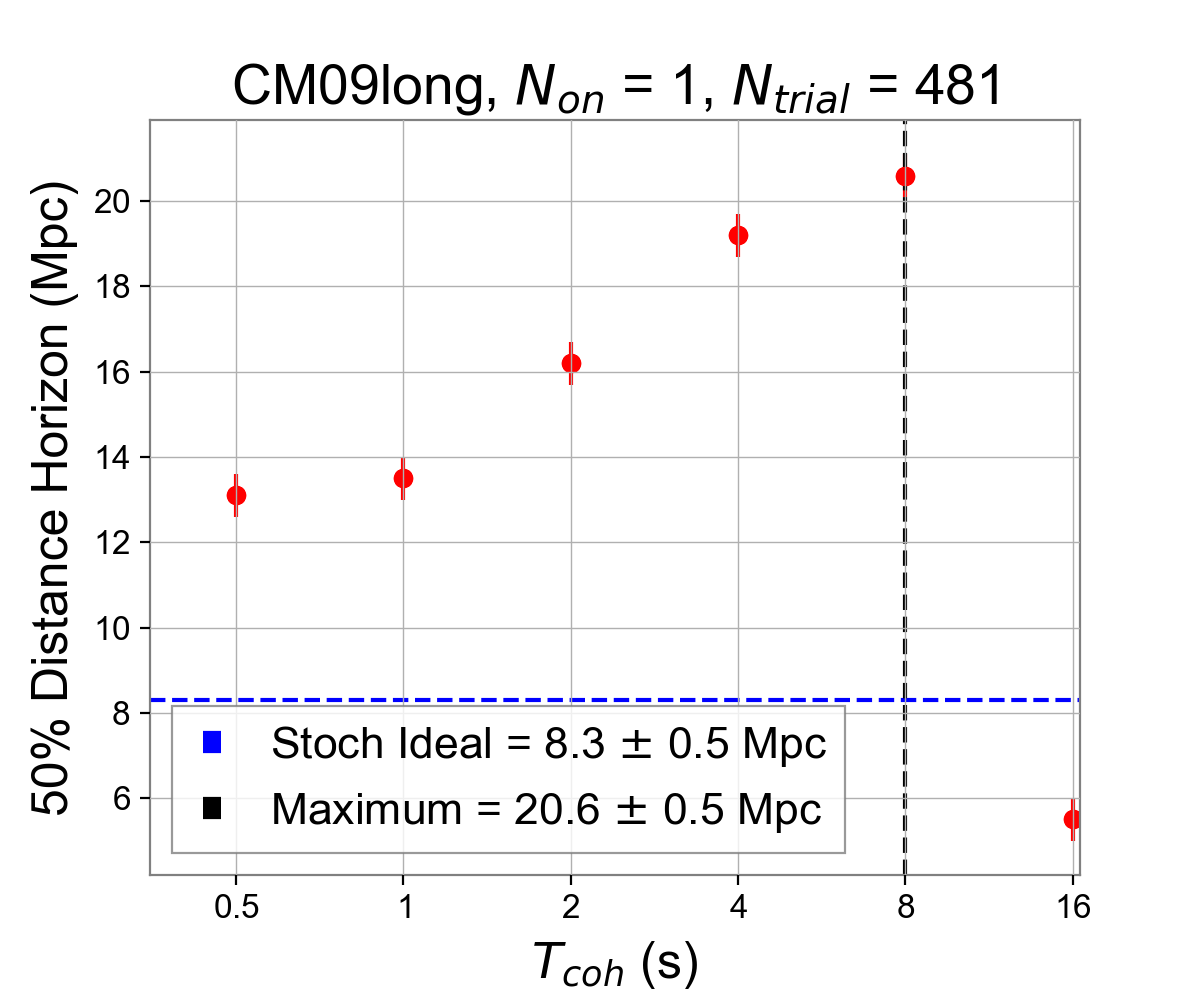}
\includegraphics[width=8.5cm]{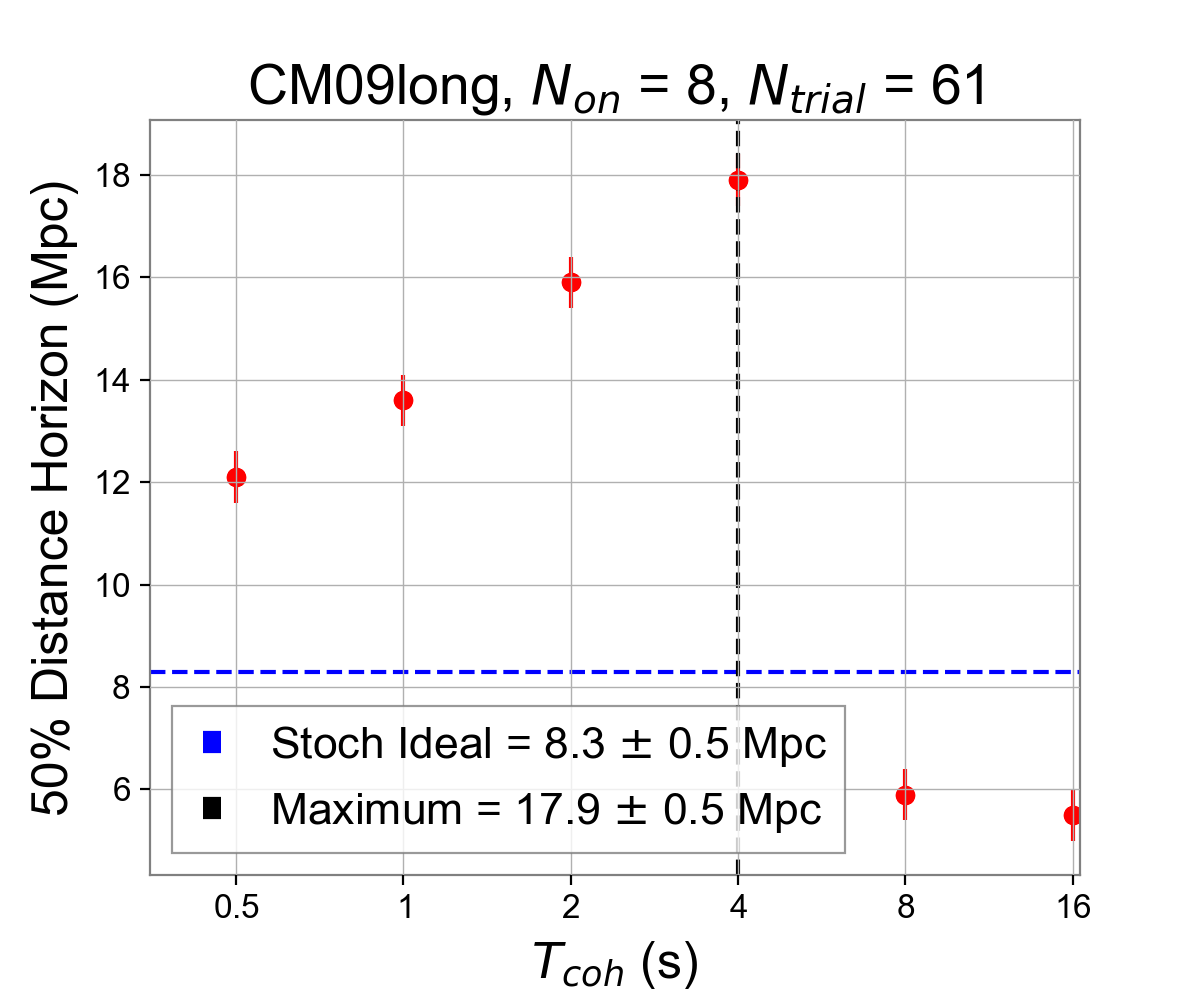}}
\hbox{
\includegraphics[width=8.5cm]{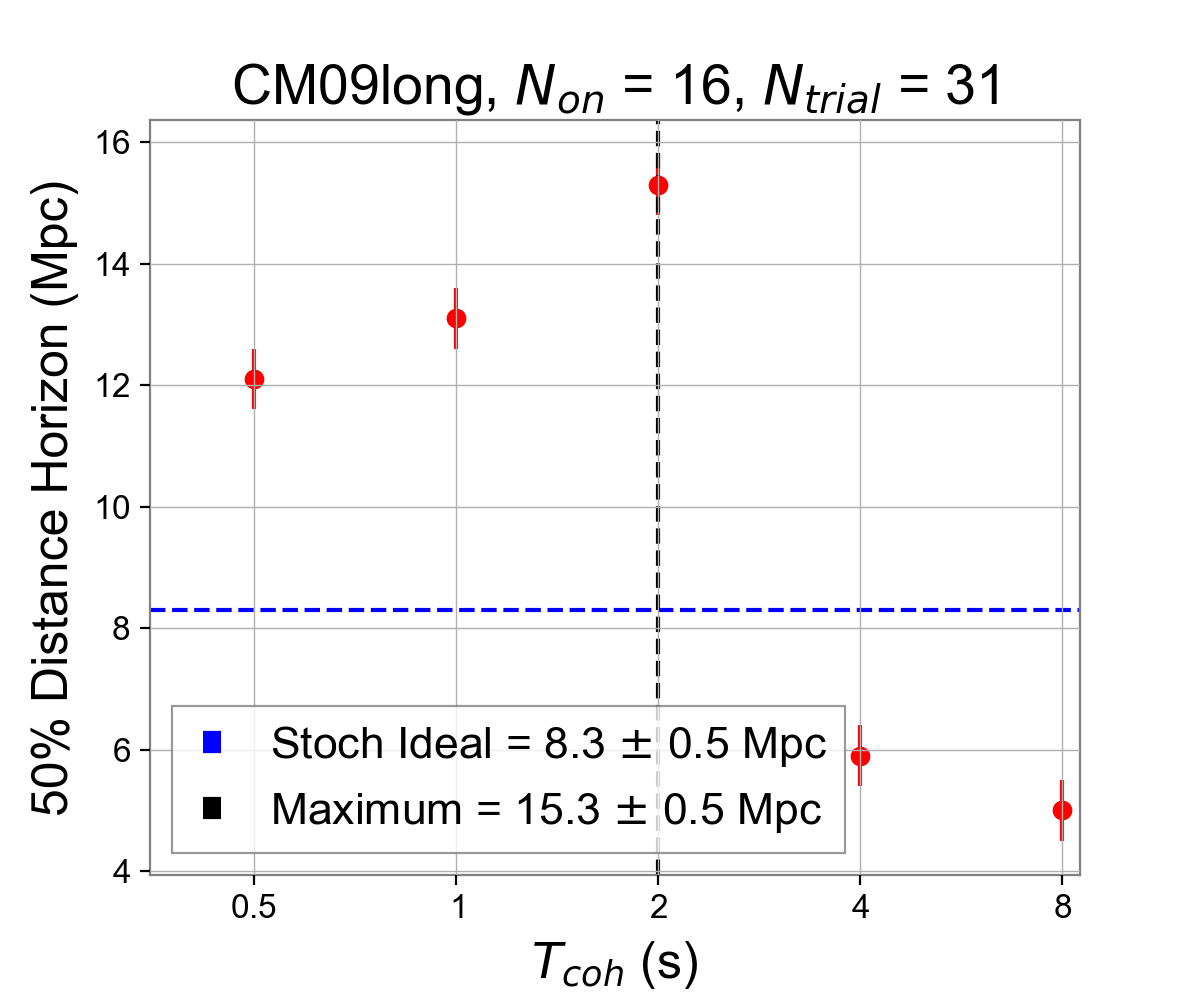}
\includegraphics[width=8.5cm]{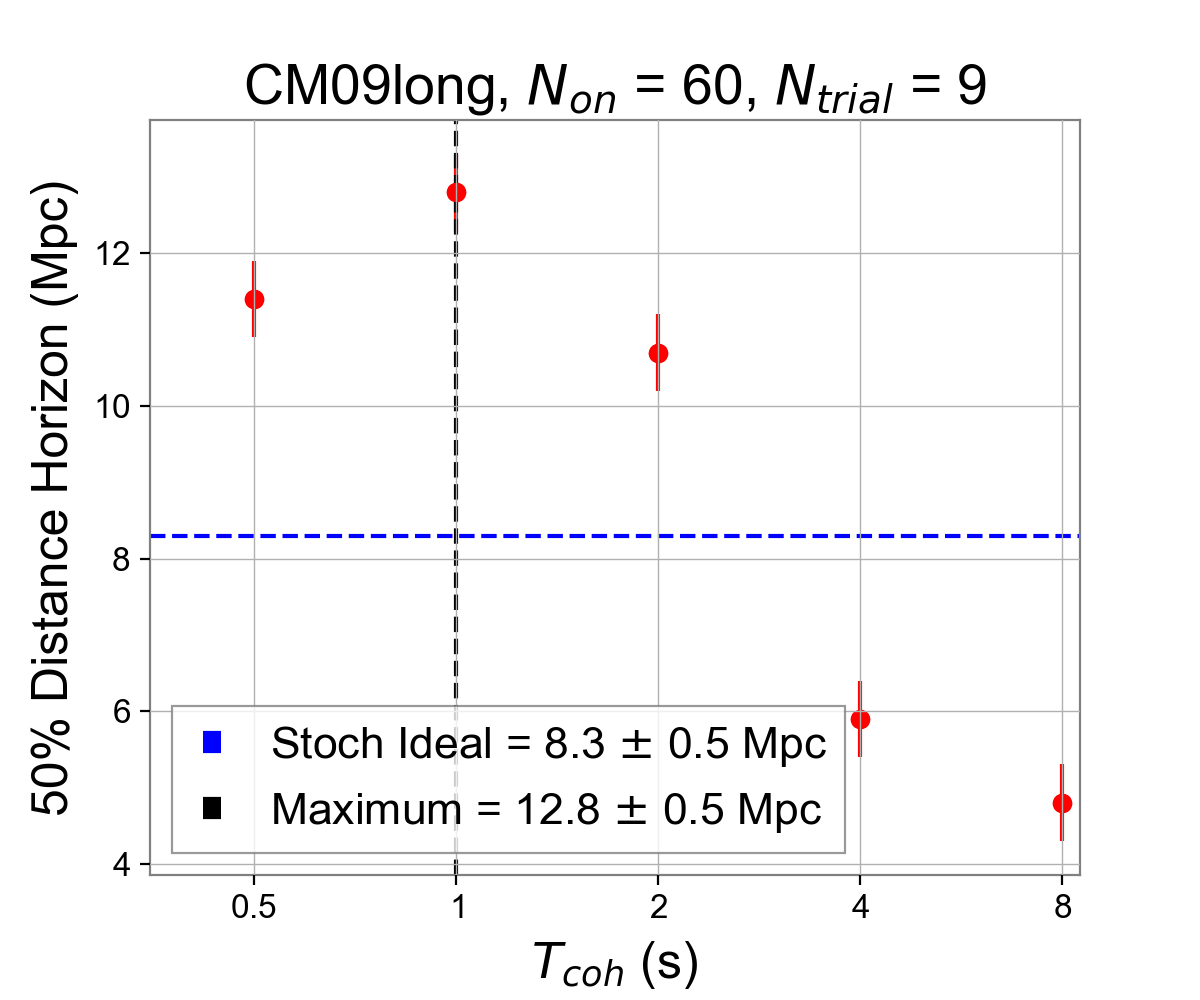}}
\hbox{
\includegraphics[width=8.5cm]{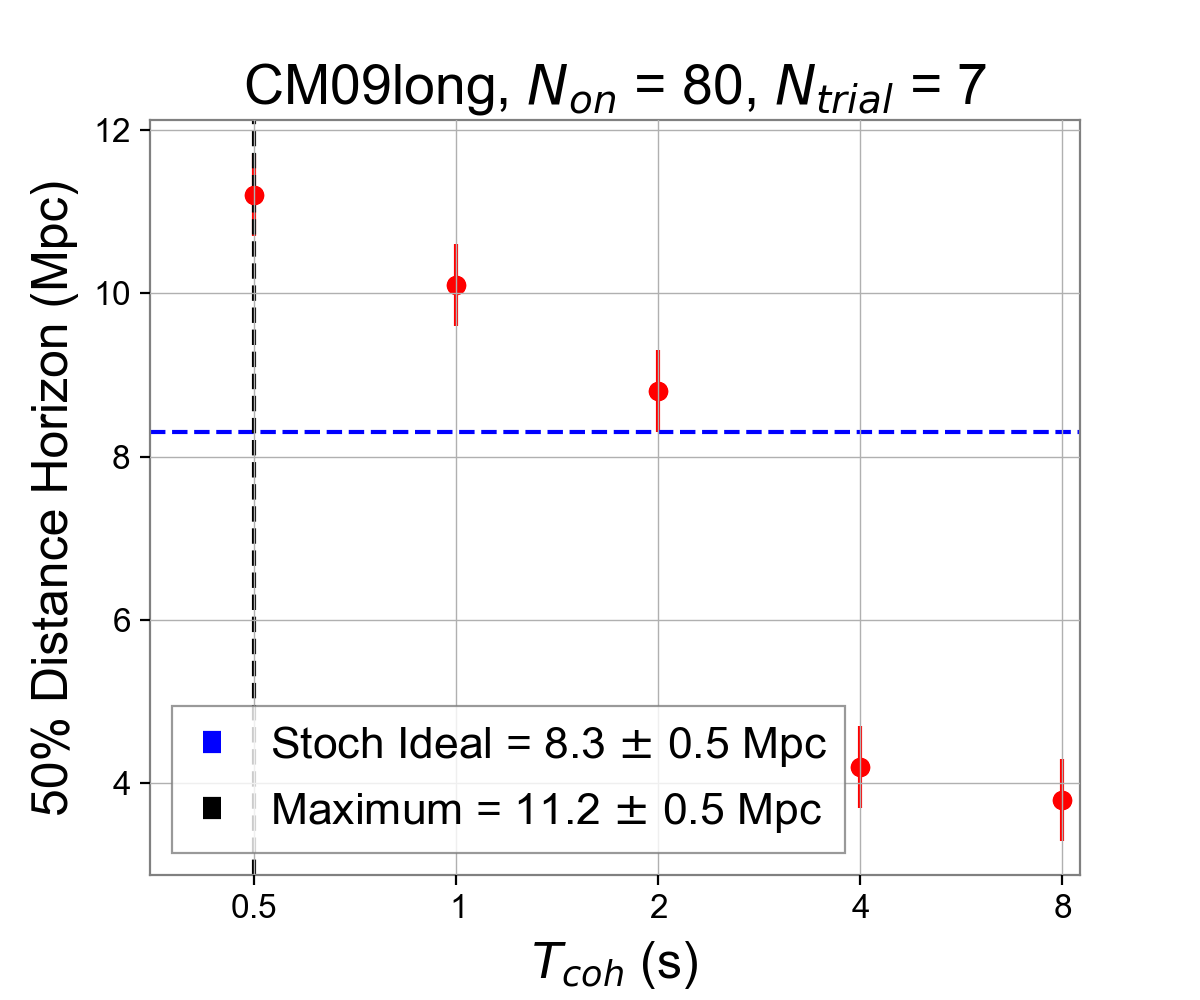}
\includegraphics[width=8.5cm]{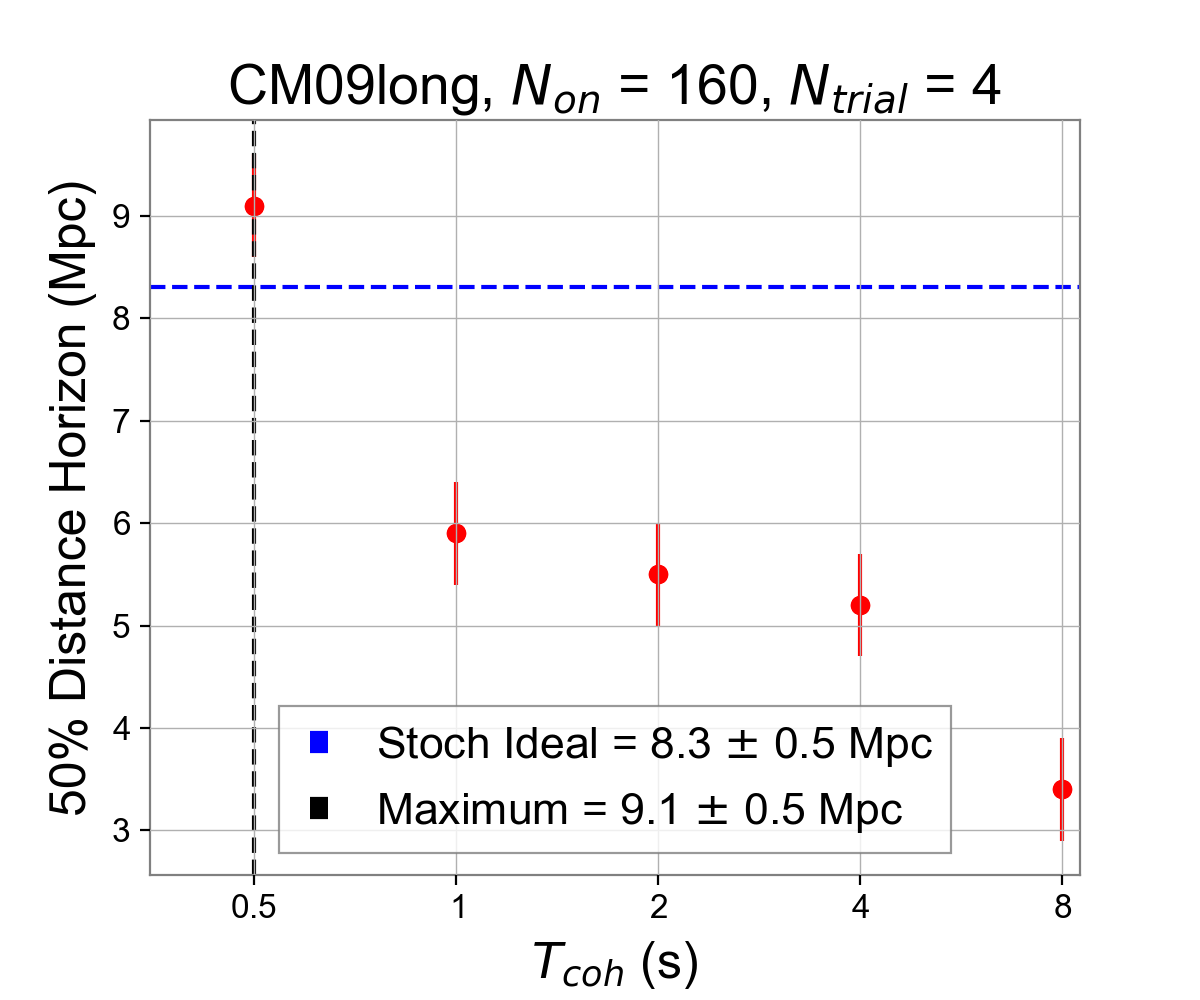}}
\caption{\label{fig:Multi-Trial}Horizon distances at 50$\%$ $FDP$ and $FAP$ of 1$\%$ for a source located at the sky position of GW 170817, and for a search of CM09long with $t_{\rm unc}=120$\,s and different values of $N_{\rm{on}}$. The computational time of a search scales with $N_{\rm{trials}}$. CoCoA distance horizons are compared with those of a single-trial stochastic search on an perfectly matching waveform (no temporal or physical uncertainties; blue-dashed lines).}
\end{figure*}

\begin{figure*}
\centering
\textbf{$t_{\rm{unc}}$ = 120\,s }\par\medskip
\hbox{
\includegraphics[width=8.5cm]{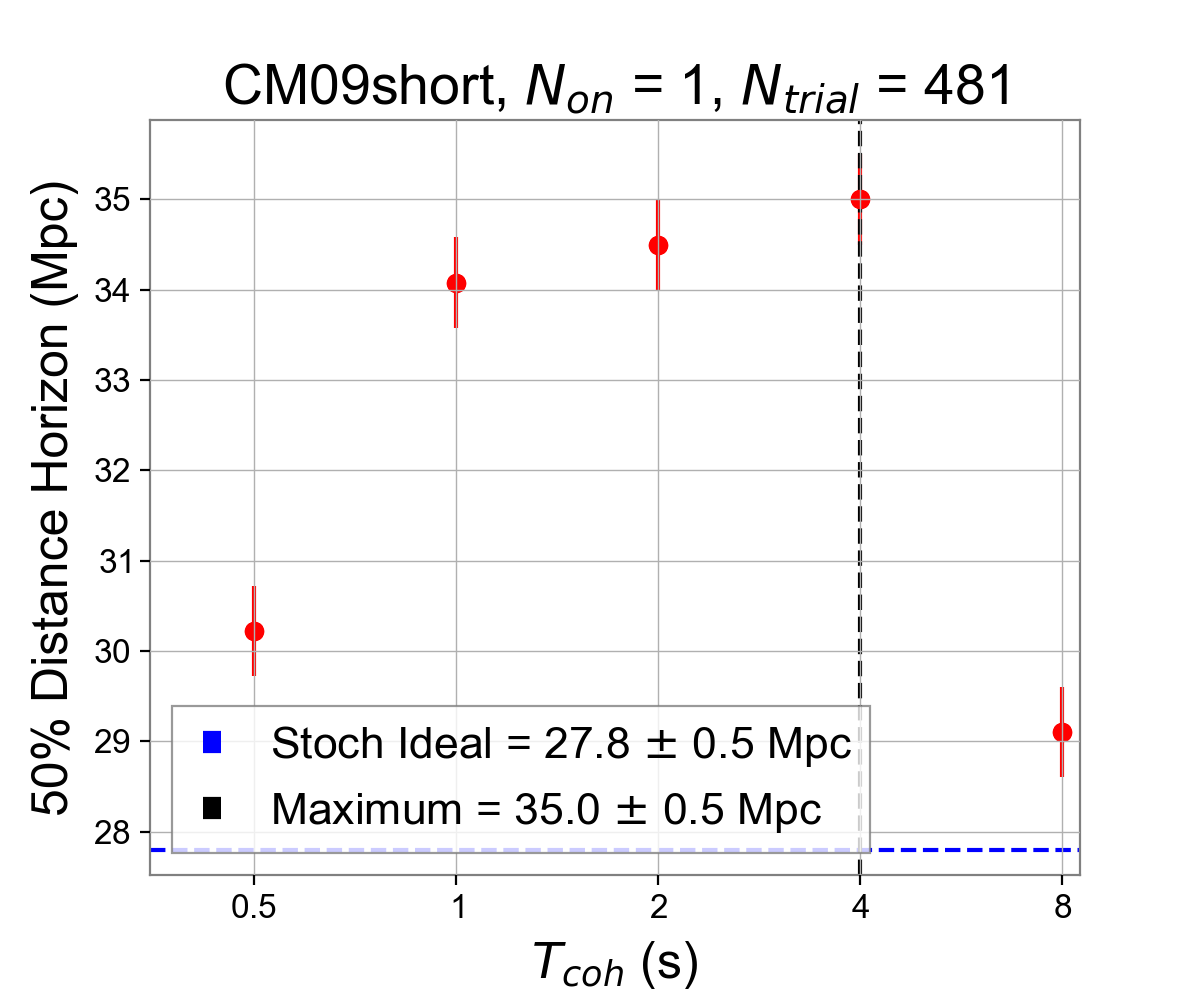}
\includegraphics[width=8.5cm]{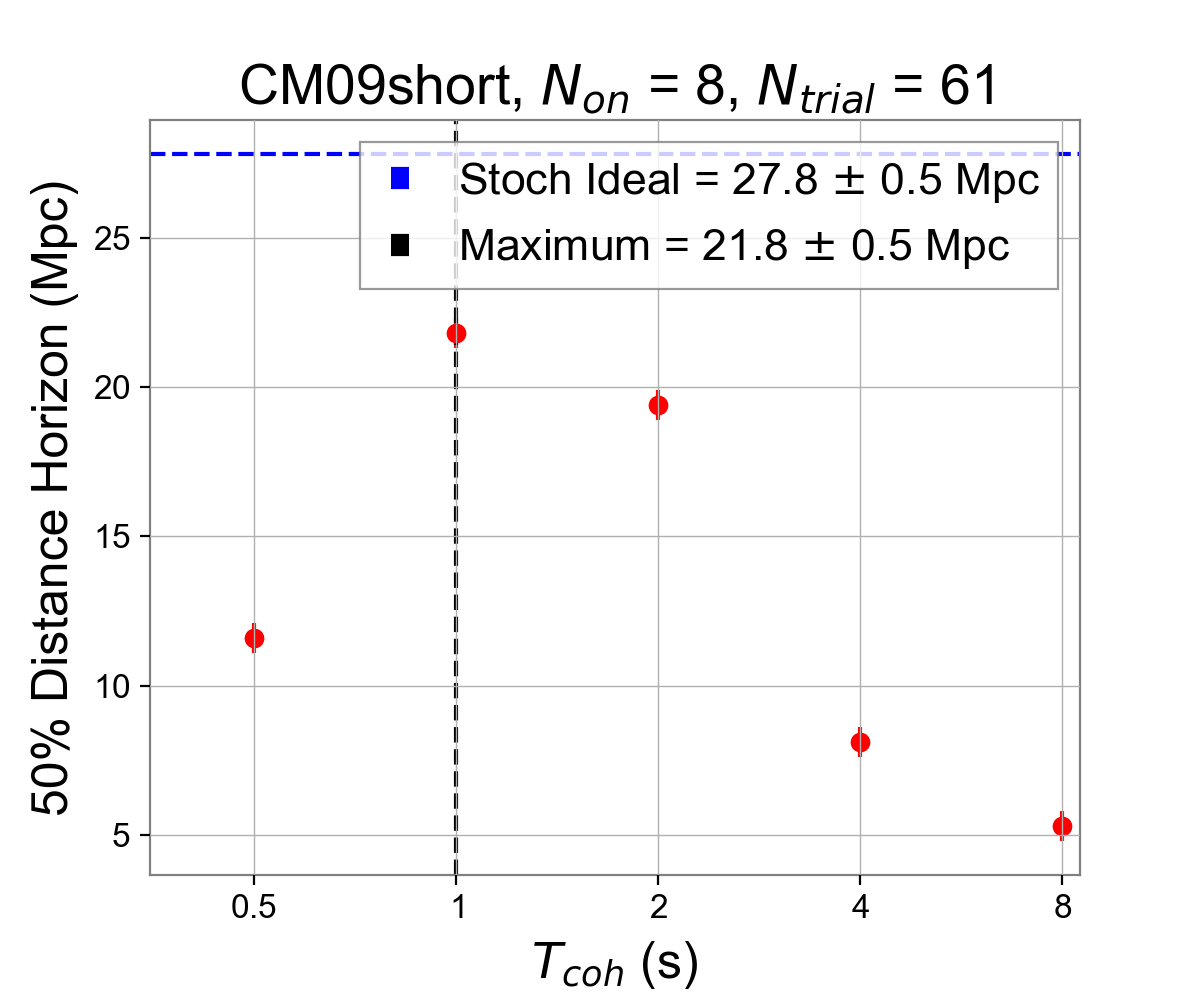}}
\caption{\label{fig:CM09shortMulti-Trial}Horizon distances at 50$\%$ $FDP$ and $FAP$ of 1$\%$ for a source located at the sky position of GW 170817, and for a search of CM09short with $t_{\rm unc}=120$\,s and different values of $N_{\rm{on}}$. The computational time of a search scales with $N_{\rm{trials}}$. CoCoA distance horizons are compared with those of a single-trial stochastic search on an perfectly matching waveform (no temporal or physical uncertainties; blue-dashed lines).}
\end{figure*}

\begin{figure*}
\centering
\textbf{$t_{\rm{unc}}$ = 2\,s }\par\medskip
\hbox{
\includegraphics[width=8.5cm]{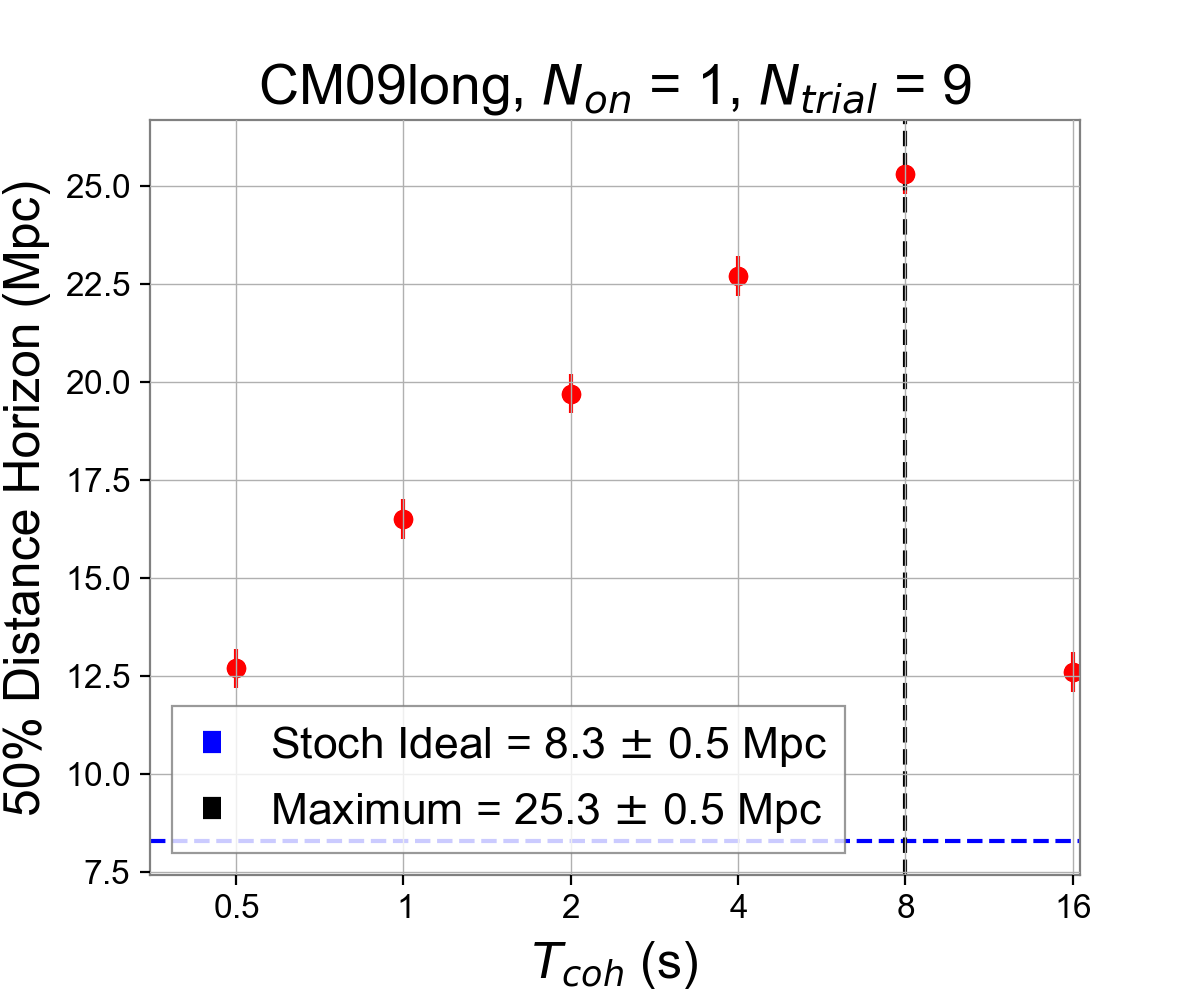}
\includegraphics[width=8.5cm]{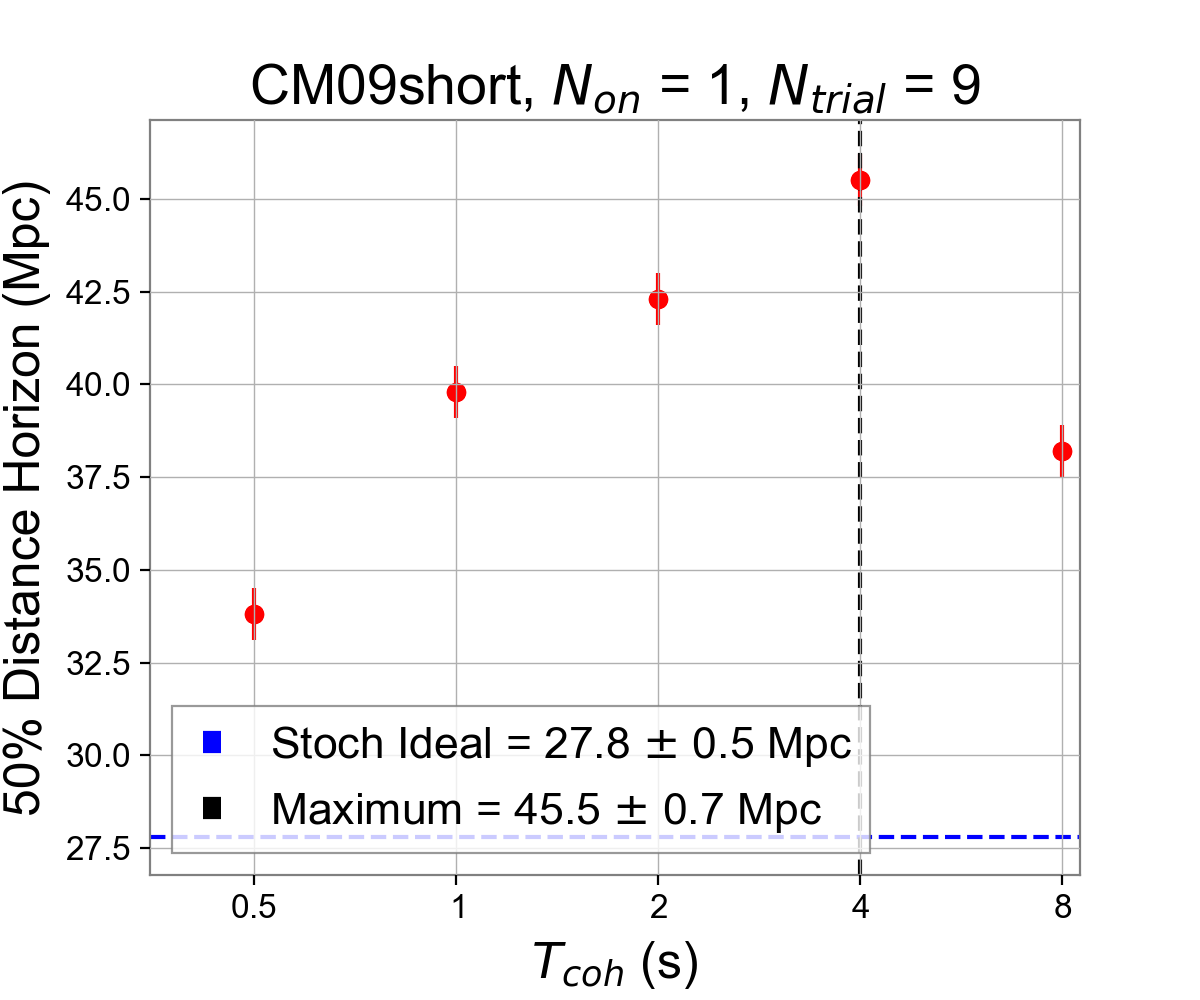}}
\caption{\label{fig:2secMulti-Trial}Horizon distances at 50$\%$ $FDP$ and $FAP$ of 1$\%$ for a source located at the sky position of GW 170817, and for searches of CM09long (left) and CM09short (right) with $t_{\rm unc}=2$\,s and $N_{\rm{on}}=1$. CoCoA distance horizons are compared with those of a single-trial stochastic search on an perfectly matching waveform (no temporal or physical uncertainties; blue-dashed lines).}
\end{figure*}
Our goal with CoCoA is to use its tunability so that we can maximize detection efficiency and ensure that the achieved distance horizon for a semi-coherent multi-trial search is always larger than even the most sensitive stochastic  (and thus less computationally expensive) search, i.e. a single-trial stochastic search with a template perfectly matching the injected waveform. This justifies the use of CoCoA over less computationally demanding stochastic algorithms \cite{Thrane2011,Thrane2013,Thrane2014,Thrane2015,Coughlin2011, 2017Postmerger}. In what follows, we demonstrate that we can reach this goal in a multi-trial CoCoA search accounting for timing uncertainties.

In Figures \ref{fig:Multi-Trial}-\ref{fig:2secMulti-Trial} we quantify the sensitivity of a CoCoA search incorporating timing uncertainties in the presence of CM09long/CM09short signals for a source located at the GW170817 position (see Section \ref{sec:bank}).  Specifically, in the various panels of Figures \ref{fig:Multi-Trial}-\ref{fig:2secMulti-Trial} we show the distance horizon corresponding to a $FDP$ of 50\% as a function of the coherence time $T_{\rm{coh}}$ of the search, for the CM09long/CM09short waveforms with different values of timing uncertainties, $t_{\rm unc}=2-120$\,s. 

Unsurprisingly, the best sensitivities (largest distance horizons) are achieved when $N_{\rm{on}}$ is equal to a single SFT baseline (as this implies minimizing the difference between the injected waveform and the closest template in the bank). Note also that the smaller the $N_{\rm{on}}$, the larger the optimal coherence time of the search. This is to be expected as larger coherent times improve sensitivity at the expense of robustness against signal uncertainties. Thus, we can afford larger coherence times for smaller differences between the closest template in our bank and the injected waveform, i.e. for smaller $N_{\rm{on}}$. These Figures also show that a coarser choice of $N_{\rm{on}}$ reduces the computational cost of the search, as larger $N_{\rm{on}}$ correspond to smaller $N_{\rm{trials}}$. This occurs at the expense of sensitivity: indeed, for $N_{\rm{on}}\approx 160$, the CoCoA distance horizon with optimized coherence time approaches the stochastic-like horizon (blue-dashed line). On the other hand, smaller $N_{\rm{on}}$ greatly improve sensitivity but imply larger number of trials and increased computational cost. To make a concrete example, 
the distance horizon we achieve for a search on CM09long with 120\,s of timing uncertainty and $N_{\rm{on}}=1$ in O2-like data is $20.6\pm0.5$\,Mpc (top-left panel in Figure \ref{fig:Multi-Trial}). However, such a search would require nearly 500 individual trials just to account for the timing uncertainty, and it would quickly become prohibitively costly computationally if one were to also account for uncertainties in the magnetar physical parameters  (see Section \ref{sec:compcost}). Thus, a more realistic search for CM09long and $t_{\rm unc}=120$\,s would be one with $N_{\rm{on}}=60$, as this produces 9 trials, which can be handled computationally even when uncertainties on the magnetar physical parameters are considered (see Section \ref{sec:compcost}). We note that an $N_{\rm{on}}=60$ CoCoA search with timing uncertainties produces a distance horizon of $12.8\pm0.5$\,Mpc. The last, rescaled for an optimally located source and for advanced LIGO nominal sensitivity (as in \cite{2018aLIGO}), corresponds to  $\approx 29$\,Mpc (only slightly less than the actual distance of GW170817).

If the timing uncertainty can be reduced to $t_{\rm unc}=2$\,s (see Section \ref{sec:timingunc}), as was the case for GW170817, then a search with $N_{\rm{on}}=1$ produces only 9 trials, and would be computationally accessible even considering uncertainties on the post-GRB magnetar properties (see discussion in Section \ref{sec:compcost}). We stress that the horizon distance of a search with $t_{\rm unc}=2$\,s and $N_{\rm{on}}=1$ is $25.3\pm 0.5$\,Mpc (see Figure \ref{fig:2secMulti-Trial}), or $\approx 57$\,Mpc for an optimally oriented source and for advanced LIGO nominal sensitivity. We note that this is comparable to the sensitivity of a single-trial search of CM09long with advanced LIGO nominal sensitivity and the same choices of $T_{\rm{coh}}$ and $T_{\rm{SFT}}$, which produces a distance horizon of $\approx$ 63\,Mpc with a $FAP$ of 1$\%$. 

Finally, as shown in Figure \ref{fig:CM09shortMulti-Trial}, faster evolving waveforms (such as CM09short) with large timing uncertainties ($t_{\rm unc}=120$\,s) are more effectively searched for with a stochastic-like algorithm rather than with a semi-coherent CoCoA approach as the last produces horizon distances smaller than the stochastic-like horizon (blue-dashed lines) for $N_{\rm{on}}\gtrsim 1$. 
However, as evident from Figure \ref{fig:2secMulti-Trial}, when the timing uncertainty can be reduced $t_{\rm unc}=2$\,s (as for GW170817), CoCoA can achieve large distance horizons ($\approx 45$\,Mpc) for a very reasonable number of trials. This implies that a search for a CM09short waveform for an optimally oriented source with advanced LIGO at nominal sensitivity could reach distances of order 100\,Mpc. We note that this is comparable to the sensitivity of a single-trial search of CM09short with advanced LIGO nominal sensitivity and the same choices of $T_{\rm{coh}}$ and $T_{\rm{SFT}}$, which produces a distance horizon of $\approx$ 140\,Mpc with a $FAP$ of 1$\%$. 

\subsection{Uncertainties in both timing and magnetar properties}
\label{sec:all-uncertainties}

In this Section we follow an approach similar to what is described in the previous one to quantify the CoCoA sensitivity and detection efficiency in the presence of both timing uncertainties and uncertainties in the physical parameters of the GRB remnant (see Section \ref{sec:timingunc}). Namely, we inject CM09long/CM09short in simulated data with sensitivity matched to LIGO O2, and run a search using a template bank that accounts for both $t_{\rm unc}$ and  uncertainties on ($\beta$, $M$, $R$, $B$). The last are taken into account by constructing a template bank where waveforms corresponding to steps of sizes $\Delta B = 10^{12}$\, G, $\Delta R = 0.02$\, km, $\Delta M = 5 \times 10 ^{-2} M_{\odot}$ around the values of CM09long/CM09short are used (see also \cite{2016Coyne}). All combinations of shifts to $M$, $R$, and $B$ are included in our template bank, giving a total of 26 unique time-frequency tracks per each of the CM09long and CM09short waveforms. We note that we do not include the exact injected waveform in our template bank so as to derive a conservative estimate of the detection efficiency.

\begin{figure*}
\centering
\hbox{
\includegraphics[width=8.5cm]{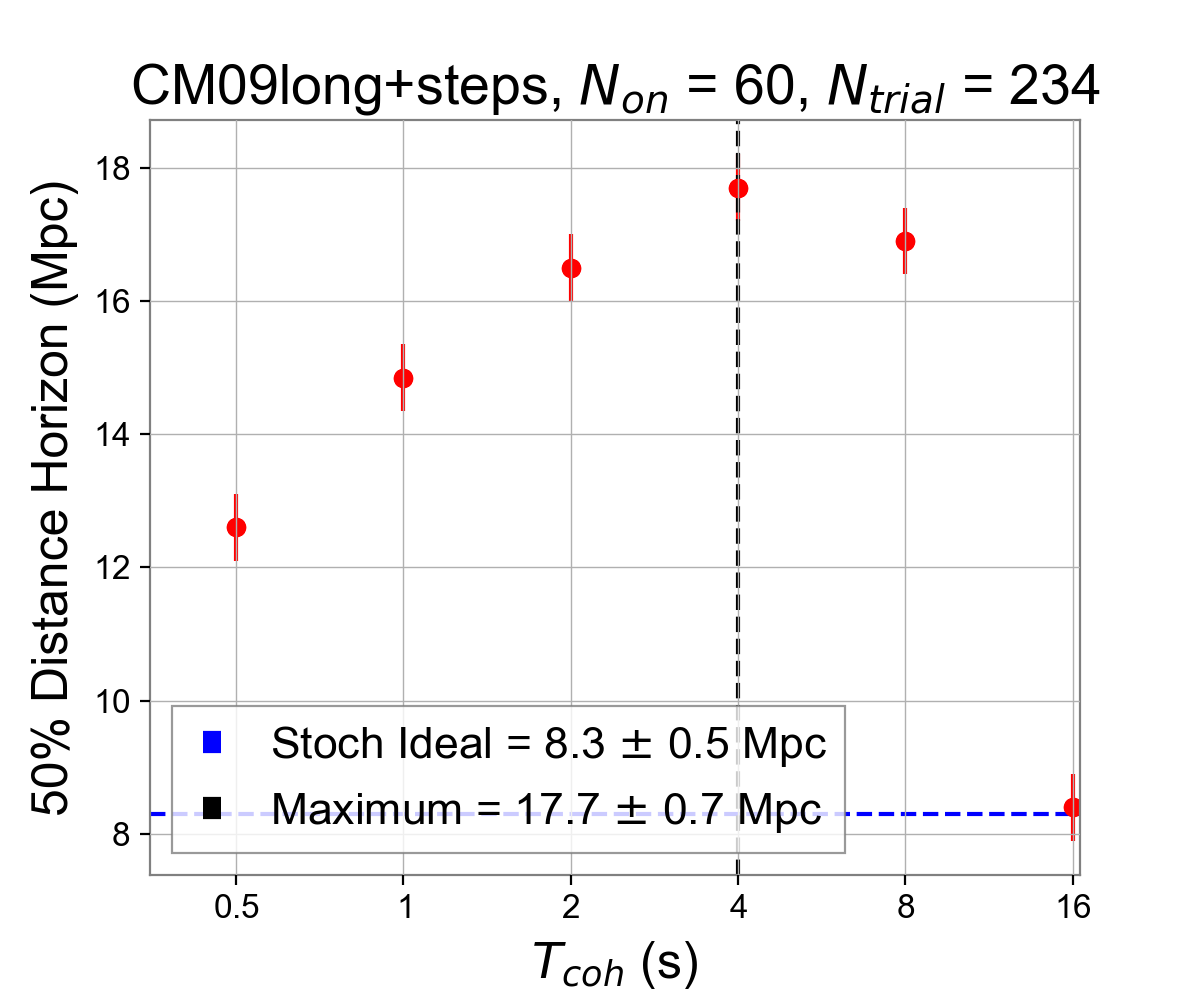}
\includegraphics[width=8.5cm]{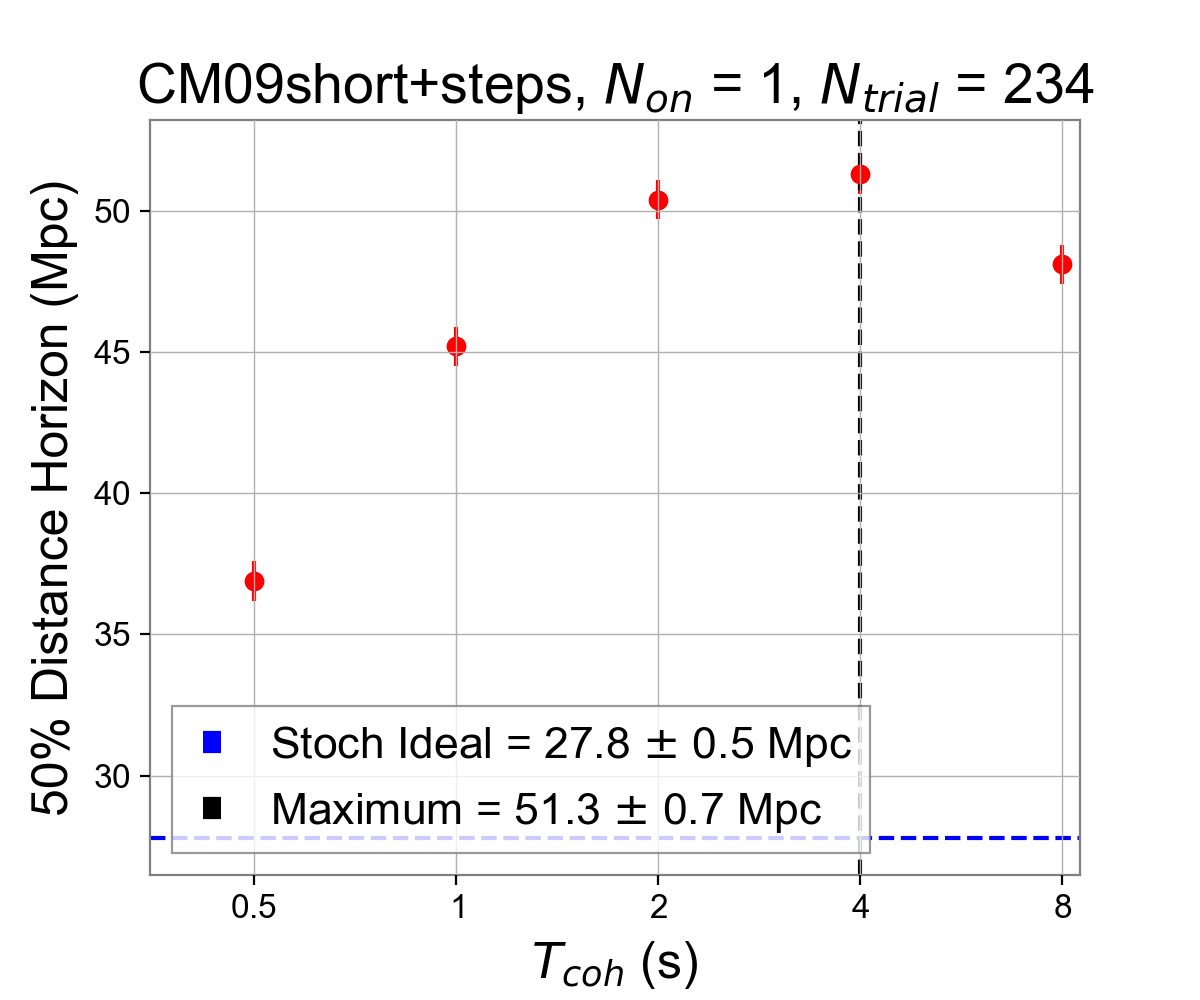}}
\caption{\label{fig:TemplateBank} Horizon distances at 50$\%$ $FDP$ and 1$\%$ $FAP$, and sky position of GW 170817 for a search of 26 sets of magnetar physical parameters obtained by shifting of $\Delta B = 10^{12}$\,G , $\Delta M = 5 \times 10^{-2}$\,M$_{\odot}$ , and $\Delta R = 0.2$\,km, the values of ($B$, $M$, $R$) for CM09long/CM09short. The searches assumes $t_{\rm unce}=120$\,s and $N_{\rm{on}}=60$ for CM09long (left), and $t_{\rm unc}=2$\,s and $N_{\rm{on}}=1$ for CM09short (right), giving 9 trials for each choice of physical parameters and 234 total trials. CoCoA distance horizons are compared with those of a single-trial stochastic search on an perfectly matching waveform (no temporal or physical uncertainties; blue-dashed lines).}
\end{figure*}

In Figure \ref{fig:TemplateBank} (left) we show the results of a search for CM09long with $N_{\rm{on}}=60$ and $t_{\rm unc}=120$\,s, which produces 9 trials accounting for timing uncertainties per each of the 26 possible choices of steps in $M$, $R$, and $B$ accounting for uncertainties in these parameters. This yields a total of 234 trials. As evident by comparing the results in Figure  \ref{fig:TemplateBank} (left panel) with those shown in the center-right panel of Figure \ref{fig:Multi-Trial}, in spite of the increased number of trials, when all possible uncertainties are considered overall the template in the bank closest to the injected waveform has a smaller mismatch than it would have by only considering timing uncertainties. In other words, small shifts in magnetar parameter values can compensate the mismatch introduced by timing uncertainties. 

Similar results are found for a search of the CM09short waveform, with $t_{\rm unc}=2$\,s, $N_{\rm{on}}=1$ (compare the right panel in Figure \ref{fig:TemplateBank} to the right panel in Figure \ref{fig:2secMulti-Trial}). In this case we find that for a fast-evolving waveform such as CM09short, small shifts in magnetar parameters combined with small shifts in the start time of GW emission may still compensate eachother, potential error of the onset time of GW emission is smaller than a single SFT. This is a surprising, yet welcome result as this compensation provides an even higher degree of sensitivity to our search. Indeed, this result provides a distance horizon of 51.3 $\pm$ 0.7 Mpc, which scales above 110 Mpc for an optimally oriented source with advanced LIGO nominal sensitivity.

\begin{figure}[ht]
\centering
\includegraphics[width=\linewidth]{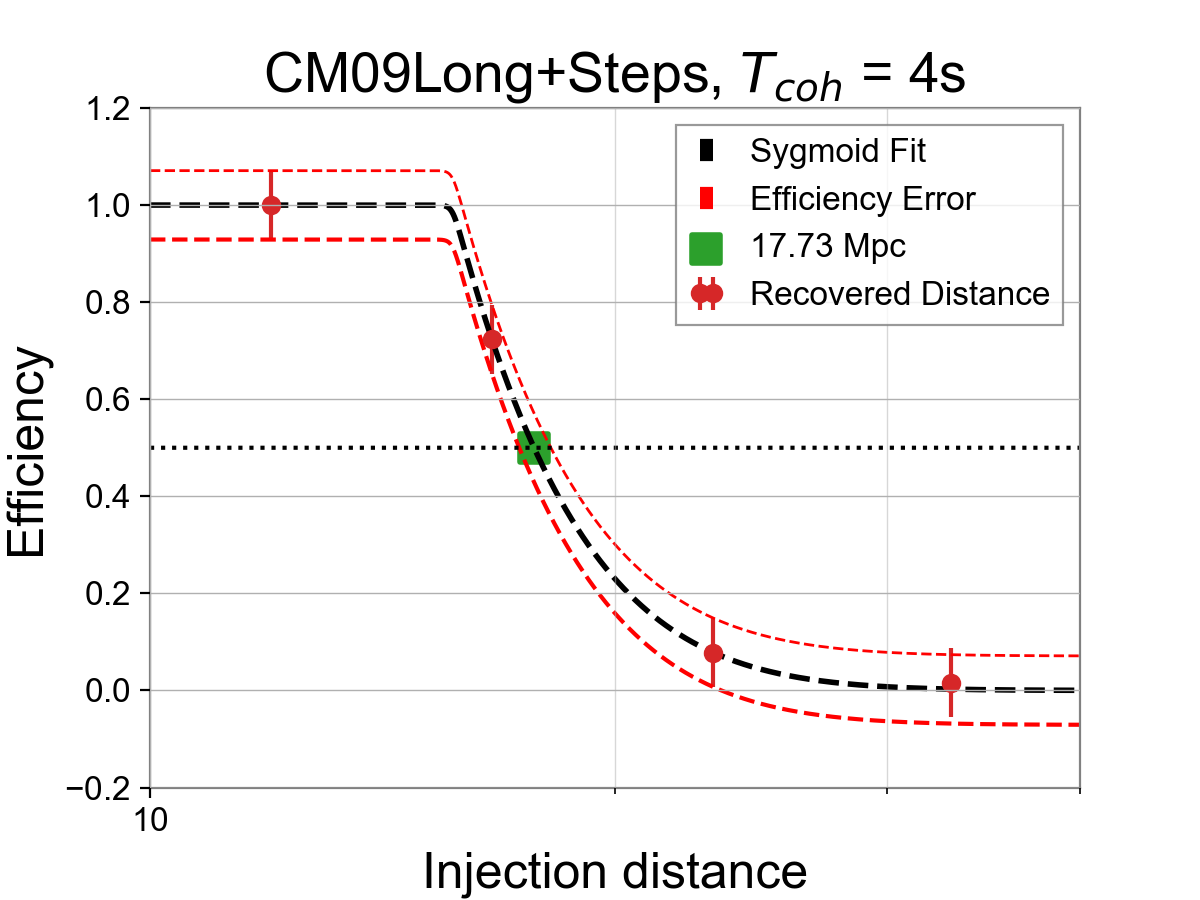}
\caption{Example of detection efficiency $(1-FDP)$ vs distance $d$ for a semi-coherent search of waveforms with small changes in physical parameters to CM09long when CM09long is injected as in Section \ref{sec:all-uncertainties}. In this example, $(N_{\rm{on}} = 60$, and $T_{\rm{coh}} = 4$\,s which is the coherence time that produces the largest distance found from a search of CM09long as in figure \ref{fig:TemplateBank}. The distance corresponding to a 50\% $FDP$ is marked in green. Error in distance is taken by first finding the error in efficiency through the DKW inequality. Efficiency errors are then connected through the sygmoid function to find errors in distance at the chosen $FDP$ (50\% in this example). See \ref{sec:multi-trial-ton-test} or \cite{2016Coyne} for more discussion.}
\label{fig:sygmoid}
\end{figure}

\section{Summary and conclusion}
\label{sec:conclusion}
In this work we have demonstrated the potential that CoCoA has for realistic targeted searches of GW signals of durations ranging from a few hundred to a few thousand seconds. Results have been shown  specifically for the case of bar-mode instabilities of millisecond magnetars formed in GRBs \cite{Corsi2009}, but can be easily generalized to other time-frequency tracks of similar durations associated with quasi-monochromatic GW signals. 

Compared to the results originally presented in \cite{2016Coyne}, we have further developed CoCoA to ensure it can run on  real GW detectors data, and that it can incorporate a multi-trial statistic allowing for searches spanning a bank of templates accounting for signal uncertainties. We have also provided order-of-magnitude estimates for the computational cost associated with various types of CoCoA searches. 

Overall our results are encouraging, as the expected distance horizons for CoCoA searches on an optimally oriented source are comparable to, or exceed, the distance of GW170817 when assuming advanced LIGO nominal sensitivity. For a binary NS merger rate in the range (0.32 - 4.760)$\times10^{-6}$ Mpc$^{-3}$\,yr$^{-1}$ \cite{2018ObservingPlan,2017Abott}, we expect 0.1 - 1 events\,yr$^{-1}$ within 40\,Mpc, and 1 - 20 events\,yr$^{-1}$ within 100\,Mpc (which should be within CoCoA reach once advanced LIGO reaches nominal sensitivity, as demonstrated here). Of these, based on current limited estimates of short GRBs opening angles (e.g., \citep{Fong2012}), $\lesssim 10\%$ would launch jets aligned with our line of sight and could thus show X-ray plateaus which would enable us to set even more stringent constraints on a potential magnetar remnant. Thus, a targeted CoCoA search for short GRB remnants that employs a full parameter space at full advanced detectors' sensitivity  may be capable of either making  detections, or else significantly constraining the most optimistic theoretical models. 

In terms of sensitivity, our results improve substantially on the ones previously presented e.g. in \cite{2017Postmerger}, but require stricter conditions on the timing uncertainties $t_{\rm unc}$ so as to ensure that a template-based CoCoA search is computationally feasible. In Appendix \ref{Sec: Compare} we discuss in more details how the CoCoA results presented here complement past searches such as the ones in \cite{2017Postmerger}.  

We finally note that magnetars may also be formed in long-duration GRBs. Thus, long GRBs (and specifically those with the characteristic X-ray plateau), will also provide interesting targets for CoCoA. Long-duration GRBs are estimated to have observed rates in the range 0.7-10$^3$\,Gpc$^{-3}$ yr$^{-1}$ (depending on luminosity; see e.g. \citep{Liang2007,Virgili2009,2015GRBEventRates}). Using the nominal advanced LIGO horizon distance for a CoCoA search of CM09long of (30\,Mpc; see Section \ref{sec:multi-trial-ton-test}), we can expect $\lesssim 0.1$\,events\,yr$^{-1}$. Thus, targeted searches for magnetars formed in long GRBs will likely need to wait for second or third generation ground-based detectors \cite{DawnIV-report, DawnIII-report, 2015Miller}.  For example, the recently funded upgrade for advanced LIGO envisions an increase in the volume of space the observatory can survey by as much as seven times \cite{DawnIV-report}, which would make long GRB searches with CoCoA come into reach on more reasonable timescales.

\appendix

\section{Effective number of trials and detection efficiency error estimation}
\label{Sec: EffTrials}

In Section \ref{sec:multi-trial-ton-test} and Figure \ref{fig:Background} we discuss results of the CoCoA background distribution from multi-trial tests. These results show a difference in the expected probability distribution (as in Eq. (\ref{eq:5.6})) and the recovered distribution. This difference is caused by overlapping time-frequency tracks of different templates in the template bank, and is quantified in Figure \ref{fig:Background} via the definition of an effective number of trials. The method we use to calculate the effective number of trials is defined below.

We start by solving Equation (\ref{eq:5.6}) for many different probability distribution functions (P($\tilde{\rho}_{\rm{max}}$)) using values of effective trials ($N_{\rm{eff,trials}}$) ranging from 0.1 to the true $N_{\rm{trials}}$ + 1 with steps of 0.1 trials. After generating the probability function in Equation (\ref{eq:5.6}) for each sampled value of $N_{\rm{eff,trials}}$ we take the integral of the probability function to generate the cumulative distribution function (CDF\footnote{Integration performed using the python scipy integration tool cumtrapz, which uses cumulative trapezoidal integration technique. For more information see https$://$docs.scipy.org/doc/scipy/reference/tutorial/integrate.html \cite{scipy} }). The CDF for each value of $N_{\rm{eff,trials}}$ is then compared to the empirical cumulative distribution function (ECDF\footnote{ECDFs are calculated using the python library statsmodels' ECDF function in the distributions sub-library. For more information see https$://\rm{www.statsmodels.org/stable/generated/statsmodels.distributions} \rm{ .empirical_distribution.ECDF.html}$ \cite{statsmodel}.}) from the recovered results. The comparison is done by using the coefficient of determination  ($R^2$) which is defined by:
\begin{equation}
    R^{2} = 1 - \frac{\sum\limits_{i} (y_{i}-f_{i})^{2}}{\sum\limits_{i} (y_{i}-\overline{y})^{2}},
    \label{eq:R2}
\end{equation}
where $y$ refers to observed data, $\overline{y}$ refers to the mean of observed data and $f$ refers to expected data \cite{R21974}. In our case $y$ refers to the ECDF from recovered results and $f$ refers to the analytic CDF generated from a given value of $N_{\rm{eff,trials}}$. The $N_{\rm{eff,trials}}$ that produces an $R^2$ value closest to 1 is taken as the chosen value of $N_{\rm{eff,trials}}$. 

The error on the effective number of trials is calculated considering that an ECDF has an error bound by the Dvoretzky-Kiefer-Wolowitz (DKW) inequality \cite{2011DKW,dvoretzky1956}. The DKW inequality is a concentration inequality which provides bounds on how a variable deviates from its expected value. Specifically, the error $\epsilon$ on the ECDF (such that the true ECDF lies between the recovered ECDF + $\epsilon$ and the recovered ECDF - $\epsilon$), is defined as: 
\begin{equation}
    \epsilon = \sqrt{\frac{\ln(\frac{2}{\alpha})}{2n}},
    \label{eq:DKW}
\end{equation}
where $1-\alpha$ is the associated probability, $n$ is the number of samples (in our case the number of background realizations as in Section \ref{sec:compcost}). 
To estimate the error on $N_{\rm{eff,trials}}$ we thus perform $R^2$ tests similarly to what described above, but using the ECDF $\pm \epsilon$. The difference between the $N_{\rm{eff,trials}}$ found when performing the $R^2$ test on the ECDF and those found when performing the same test on the ECDF $\pm \epsilon$ is taken as the error in $N_{\rm{eff,trials}}$. As the $R^2$ test only considers discretely sampled values of $N_{\rm{eff,trials}}$, 0.1 effective trials apart, we also add an additional systematic error of 0.05 on the estimated $N_{\rm{eff,trials}}$.

The DKW inequality is also used in the calculation of errors on the detection efficiency and distance horizons. Because the $FAP$ threshold for a given search makes use of the ECDF, the $\epsilon$ in Equation (\ref{eq:DKW}) also puts bounds on our error on the detection efficiency for a chosen $FAP$ (red errors bars in Figure \ref{fig:sygmoid}). The recovered efficiencies at each injected distance, as well as their upper- and lower-error ranges are then fit to sygmoid curves (see the dashed lines in Figure \ref{fig:sygmoid}). The distance that corresponds to the point where the chosen $FDP$ level (black-dotted line in Figure \ref{fig:sygmoid}) crosses the sigmoid fit to the detection efficiency (black-dashed line in Figure \ref{fig:sygmoid}), is then taken as the distance horizon for that given $FDP$ (marked in green in Figure \ref{fig:sygmoid}). The error on such distance is estimated by using the points where the sygmoids fits to the upper and lower bounds of the efficiency curve (red-dashed lines in Figure \ref{fig:sygmoid}) cross the chosen $FDP$  (black-dotted line in Figure \ref{fig:sygmoid}).

\section{Order-of-magnitude comparison with previous GW170817 post-merger results}
\label{Sec: Compare}

Searches for post-merger GWs from secularly unstable magnetars with parameters matched to those of Bar1-Bar6 (see Section \ref{sec:testwave}) have been performed for GW170817 using the Stochastic Analysis Multi-detector Pipeline (STAMP) \cite{Coughlin2011,Thrane2013,2017Postmerger}. STAMP searches for excess power in time-frequency maps by cross-correlating data streams of different detectors and using pattern recognition algorithms rather than a template bank of time-frequency tracks. For the pattern recognition, STAMP uses both seed-based (Zebraguard) and seedless (Lonetrack) algorithms. Because STAMP searches are most similar to the CoCoA stochastic limit, here we compare expectations for a stochastic-limit CoCoA search on Bar1-Bar6 with STAMP results on these same waveforms reported in \cite{2017Postmerger}.

We consider a CoCoA search similar to that described in Section \ref{sec:bank}, with $T_{\rm{SFT}} = 0.25$\,s, and $FAP$ and $FDP$ matching those used for the  STAMP search in \cite{2017Postmerger}. An appropriate template bank for such as CoCoA search would include waveforms Bar1-6, as well as a range of other waveforms with same $M$ and $\beta$ but spanning ranges of $R=12-14$\,km for the NS radius, and $B=10^{13}-5 \times 10^{14}$ G for the magnetic field, in steps of $\Delta B = 10^{12}$\, G and $\Delta R = 0.02$\, km (see Section \ref{sec:all-uncertainties}).
For this choice in step size the CoCoA template bank would contain 100 templates to span the possible values of $R$, and 490 templates to span the possible values of $B$. Temporal uncertainty can be accounted for by choosing $t_{\rm{unc}} = 2$\,s, which is comparable to the delay between the merger time of GW170817 and its associated GRB\,170817A (see Sections \ref{sec:compcost} and \ref{sec:multi-trial-ton-test} for more discussion). With $\Delta t_{\rm{on}} = 0.25$\,s (as in Section \ref{sec:bank}), this would result in 9 choices of $t_{\rm on}$ for each ($\beta$, $M$, $R$, $B$). Thus, we expect a CoCoA search to include $N_{\rm{trials}} = 100 \times 490 \times 9 = 441\times10^3$ trials.

\begin{figure}
\centering
\includegraphics[width=\linewidth]{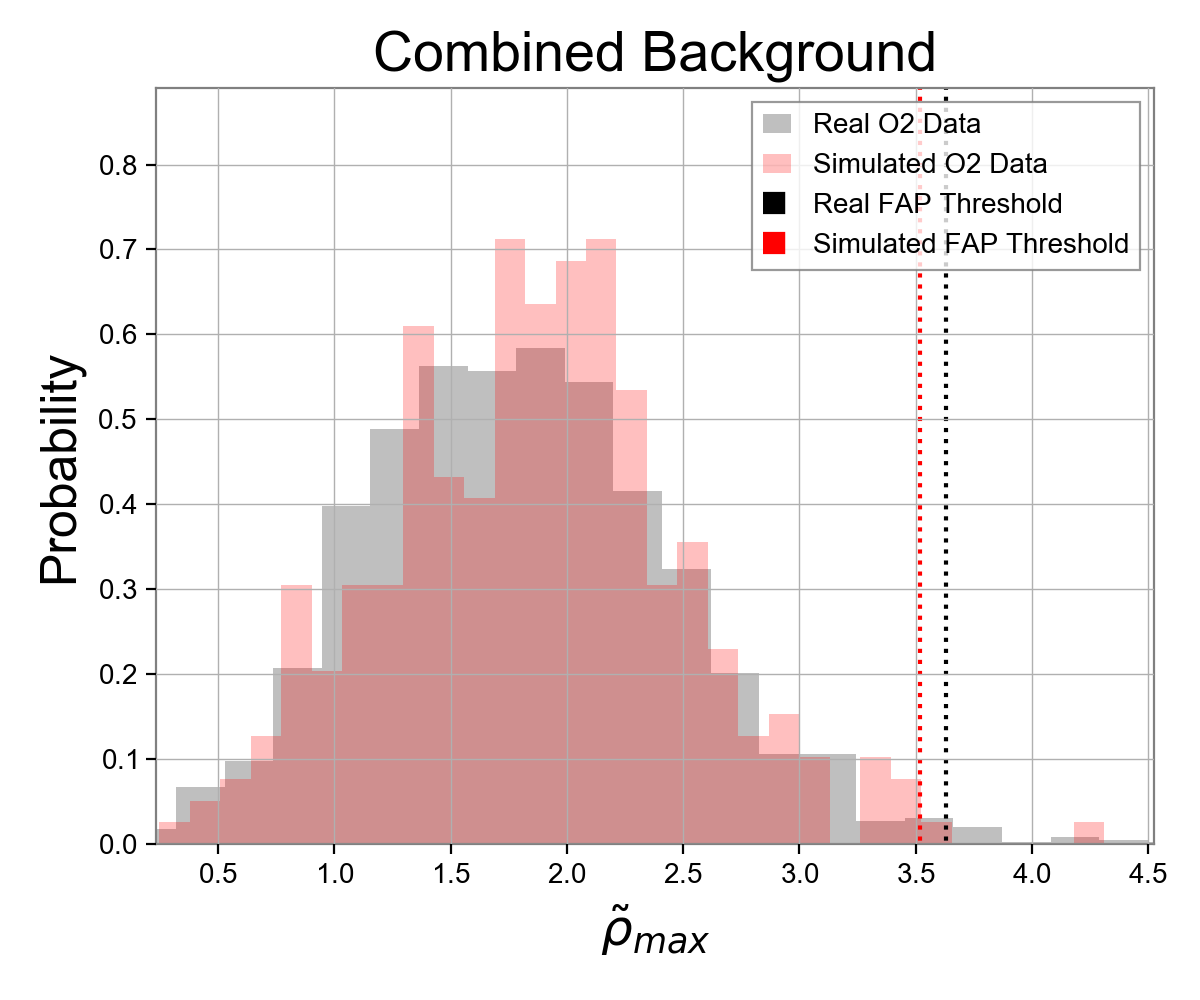}
\caption{The combined background distribution of the $\tilde{\rho}_{\rm max}$ statistic for a search on O2 data that considers Bar1-6, the waveforms with all combinations  of  $(\beta, M, R, B)$  one  step  away from  those of Bar1-6, and 9 choices of $t_{\rm{on}}$, for a total number of trials of $N_{\rm{trial}}=481$. The grey histogram shows the results of the search over 9.1 days of real O2 data before the time of the merger of GW170817. This background is used to set the FAP threshold for the comparison to STAMP test. The pink histogram shows the background when using simulated O2-like Gaussian noise similar to what is used in Section \ref{sec:multi}. The thresholds for a FAP of $1\%$ for both the real and simulated data are show using black- and red-dashed lines, respectively.}
\label{fig:compare-bkg}
\end{figure}

\begin{figure}
\centering
\includegraphics[width=\linewidth]{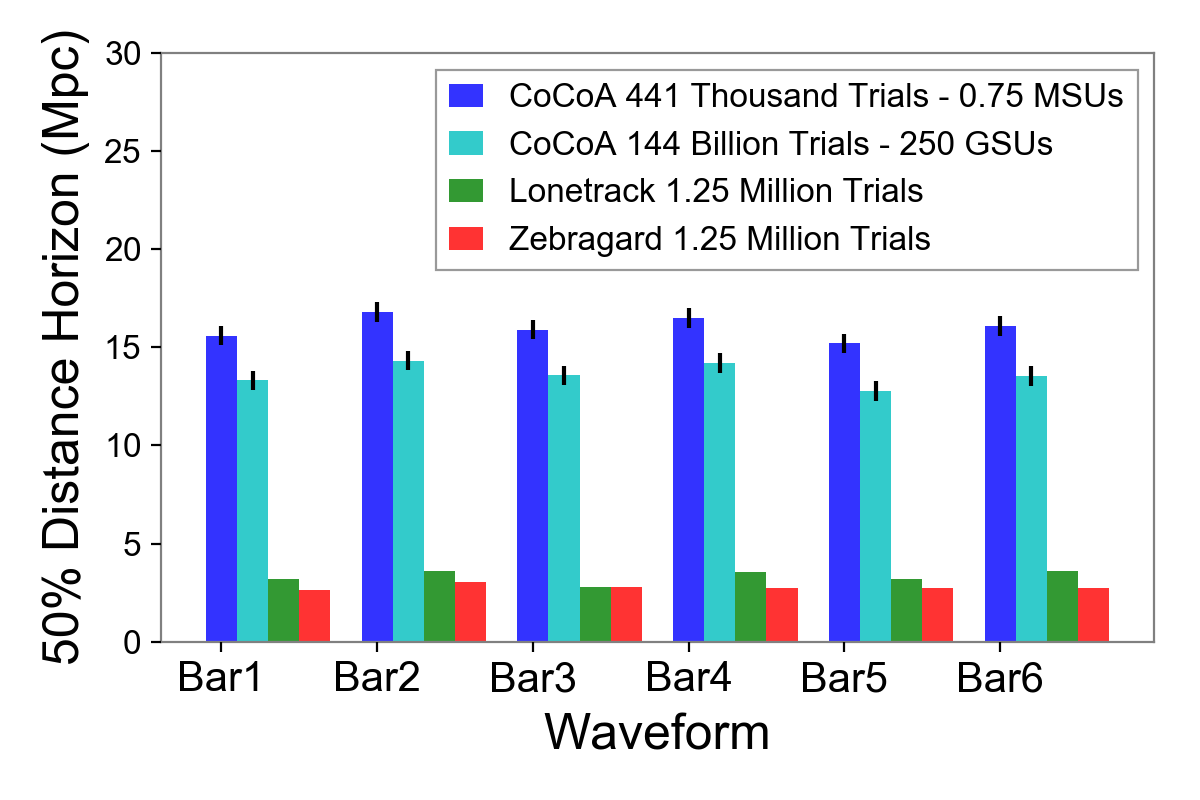}
\caption{Horizon distances for STAMP as in \cite{2017Postmerger}  compared to the ones of a CoCoA stochastic search. CoCoA (dark and light blue), even with the less sensitive stochastic limit, is more sensitive than STAMP (red and green). But, the gained sensitivity comes at the expenses of computational cost. A CoCoA search with $t_{\rm{unc}} = 8.5$\,d (light blue), like that used by the STAMP search  presented in \cite{2017Postmerger} produces a number of trials 6 orders of magnitude larger. Similarly, a CoCoA search with $t_{\rm{unc}}=2$\,s (dark blue) produces a number of trials only about a factor of two smaller than a STAMP search with a much longer timing uncertainty of  $t_{\rm{unc}}=8.5$\,d. 
We note that the Coherent Wave Burst (cWB) pipeline \cite{2016Klimenko} has also produced upper-limits for the GW170817 post-merger search that are comparable to the STAMP ones shown here (see \cite{2017Postmerger}).
\label{fig:STAMPCompare}}
\end{figure}

In order to estimate the sensitivity of a CoCoA search without spending a large amount of computational time, we compute our background statistic using a reduced template bank that only considers Bar1-6 and all combinations of physical parameters that are one step away from Bar1-6. The background is built using 9.1 days of coincident detector data during the O2 LIGO run, starting 20 days before the GW170817 merger and ending 1 hour before. The results are displayed in Figure \ref{fig:compare-bkg}. From the grey histogram in this Figure we calculate the $\tilde{\rho}_{\rm th}$ corresponding to a $FAP=1\%$ and find this to be in excellent agreement with what expected from simulated Gaussian noise with O2-like sensitivity (pink histogram in  Fig. \ref{fig:compare-bkg}). Next, we use Eq.  (\ref{eq:combined$FAP$calculated}) to estimate the $\tilde{\rho}_{\rm th}$  of a search with $N_{\rm{trials}}=441\times10^3$ (see above) and $FAP = 1\%$. We then estimate the CoCoA distance horizon for Bar1-Bar6 by injecting those signals in the longest O2 stretch of data closest to the trigger time of GW170817 (starting  at  GPS  time  1186898000), and searching with a template bank that considers all combinations of physical parameters one step away from the injected waveform (this is similar to what is done in Section \ref{sec:all-uncertainties}). The results of this test are compared to STAMP's results for GW170817 in Figure \ref{fig:STAMPCompare}. 

From Figure \ref{fig:STAMPCompare} we see that CoCoA (blue bars), even in its least sensitive stochastic limit, is more sensitive than STAMP. But, the gained sensitivity comes at the expenses of computational cost. This is ultimately related to the fact that while CoCoA is a template-based search that considers the expected physics behind the time-frequency tracks it searches for, STAMP time-frequency maps are build using analytic methods that do not consider specific models.  While this reduces the STAMP search sensitivity, it makes it computationally more feasible in the presence of large signal uncertainties. Indeed, the red and green bars in Figure \ref{fig:STAMPCompare} show the results of the STAMP search reported in \citet{2017Postmerger}. The last targeted bar-like GWs starting at the time of the GW170817 merger and ending $\sim$ 8.5 days after the merger, thus allowing for a $t_{\rm{unc}}$ much greater than the 2\,s considered for a CoCoA search (light and dark blue bars). The STAMP search was carried out using time-frequency maps of duration 500\,s, $t_{\rm on}$ times with 50$\%$ overlap from the previous time-frequency map, and an SFT duration of 1\,s, for a total of $1250\times10^3$ trials. If we were to build a CoCoA search with the same choice of $t_{\rm{unc}} \sim 8.5$\,days and keeping $\Delta t_{\rm{on}} = 0.25$\,s, we would need $\sim 3\times10^{6}$ choices of $t_{\rm on}$ for each ($\beta$, $M$, $R$, $B$) and $144\times10^{9}$ trials for the full search (Fig. \ref{fig:STAMPCompare}, light blue). A search of this magnitude would cost 250 GSUs, which is 5 orders of magnitude larger than other LIGO searches, (e.g., \cite{2019Caride}) and is therefore computationally unfeasible.

In conclusion, we can say that the STAMP and CoCoA approaches are complementary, and we advocate for running searches with both as the most likely way for maximizing chances of detecting intermediate-duration post-merger signals. 

\section{Additional Tests}
\label{App. Additional Tests}
In this Section we show how the performance of CoCoA changes by changing some of the assumptions we made in Section \ref{sec:multi-trial-ton-test}.  Specifically, we consider (i) changing the $FDP$ from 50\% to 10\%; and (ii) randomizing the injection time $t_{\rm{inj}}$ rather than having it always fall exactly in between two adjacent SFT bins (a choice that maximizes the mismatch between injections and templates). We test these changes on two representative searches. A first search of CM09long with $t_{\rm{unc}} = 120$\,s, $N_{\rm{on}} = 60$, and $T_{\rm{coh}}= 1$\,s; and a second search of CM09short with $t_{\rm{unc}} = 2$\,s, $ N_{\rm{on}} = 1$, and and $T_{\rm{coh}}= 4$\,s. These coherence time values are chosen based on the optimization procedure shown in Figures \ref{fig:Multi-Trial} and \ref{fig:2secMulti-Trial}. 

Our results are reported in Table \ref{tab:Add. Tests},  where for reference we also show the distance horizons obtained for $FDP=50\%$, for injections times matching exactly the start times of the template waveforms (which eliminates the mismatch between the two; see $d^{90\%}_{\rm nomis}$  and $d^{90\%}_{\rm nomis}$  in Table \ref{tab:Add. Tests}), and for injection times always in between adjacent SFT bins (which maximizes the mismatch between injected and template waveforms; see $d^{90\%}_{\rm maxmis}$  and $d^{90\%}_{\rm maxmis}$  in Table \ref{tab:Add. Tests}).

\begin{table*}[hbt!]
\centering
\caption{\label{tab:Add. Tests}CoCoA performance with varying $FDP$ and injection times. See Appendix \ref{App. Additional Tests} for discussion. 
} 
\begin{tabular}{cccccccccc}
\hline
\hline
Waveform & $t_{\rm{coh}}$ & $t_{\rm{unc}}$ & $N_{\rm{on}}$  & $d^{50\%}_{\rm nomis}$  & $d^{90\%}_{ \rm nomis}$ & $d^{50\%}_{\rm maxmis}$  & $d^{90\%}_{\rm  maxmis}$ & $d^{50\%}_{\rm randinj}$ & $d^{90\%}_{\rm randinj}$ \\
& (s) & (s) & & (Mpc) & (Mpc) & (Mpc) & (Mpc) & (Mpc) & (Mpc)\\
\hline
CM09long & 1 & 120 & 60 & 15.0 $\pm$ 0.5 & 12.3 $\pm$ 0.4 & 12.8 $\pm$ 0.5 & 10.4 $\pm$ 0.4 & 13.6 $\pm$ 0.5 & 11.5 $\pm$ 0.4 \\
\hline
CM09short & 4 & 2 & 1 & 60.0 $\pm$ 0.7 & 46.0 $\pm$ 0.7 & 45.5 $\pm$ 0.7 & 37.9 $\pm$ 0.6 & 52.8 $\pm$ 0.7 & 42.6 $\pm$ 0.6 \\
\hline
\end{tabular}
\end{table*}
Unsurprisingly, when the onset time is randomized, the sensitivity of the search (see $d^{50\%}_{\rm randinj}$ and $d^{90\%}_{\rm randinj}$ in Table \ref{tab:Add. Tests}) falls in between the two extremes of no mismatch and maximized mismatch. Also unsurprisingly, we find that decreasing the allowed $FDP$ reduces the sensitivity of the search, so the distance horizons for $FDP=10\%$ are $\sim 15\% - 30\%$ smaller than for $FDP=50\%$.

\acknowledgments
This work is supported by the National Science Foundation via CAREER award \#1455090 (PI: Alessandra Corsi). We thank Benjamin J. Owen for general discussions on topics related to this work. This paper has document number LIGO-P1500226.

\bibliographystyle{apsrev-titles}
\bibliography{main}
\bibstyle{apsrev-titles}

\end{document}